\documentclass[aps,prd,twocolumn,superscriptaddress,nofootinbib,preprintnumbers]{revtex4-1}
\usepackage{amsmath}
\usepackage{graphicx}
\usepackage{subfigure}
\usepackage{amssymb}
\usepackage{xcolor}
\usepackage{multirow}
\usepackage{cancel}
\usepackage{color}
\usepackage{ulem}
\usepackage{listings}
\usepackage{xcolor}
\usepackage{float}
\usepackage{slashed}
\usepackage{wrapfig}
\usepackage{mathtools}
\usepackage[colorlinks,citecolor=blue]{hyperref}
\usepackage[T1]{fontenc}  
\usepackage[utf8]{inputenc}  
\usepackage{times}
\usepackage{comment}
\usepackage{dsfont}
\usepackage{orcidlink}
\begin{document}
	
\title{Gravitational waves from supermassive right-handed neutrinos produced at preheating}

\author{Shinya Kanemura}
\email{kanemu@het.phys.sci.osaka-u.ac.jp}
\affiliation{Department of Physics, The University of Osaka, Toyonaka, Osaka 560-0043, Japan}

\author{Kunio Kaneta\orcidlink{0000-0001-5391-2204}}
\email{kaneta@ed.niigata-u.ac.jp}
\affiliation{Faculty of Education, Niigata University, Niigata 950-2181, Japan}
	
\author{Dibyendu Nanda\orcidlink{0000-0002-7768-7029}}
\email{dnanda@het.phys.sci.osaka-u.ac.jp}
\affiliation{Department of Physics, The University of Osaka, Toyonaka, Osaka 560-0043, Japan}
\affiliation{Department of Physics, School of Basic Sciences and Humanities, SR University, Warangal 506371, India}

\begin{abstract}
The post-inflationary production of supermassive particles can have profound implications for the thermal history of the universe and may leave observable imprints in the gravitational wave (GW) background. In scenarios where the inflaton couples predominantly to heavy fields, say right-handed neutrino (RHN), non-perturbative mechanisms such as parametric resonance can lead to their efficient production, even when their masses exceed the inflaton mass. Once produced, the RHNs emit gravitons through bremsstrahlung as they decay into the Standard Model (SM) particles via $N\rightarrow \ell + H$, enabled by the unavoidable minimal coupling to gravity, sourcing a stochastic GW background. We study this mechanism within the framework of $\alpha-$attractor inflationary models, highlighting how the resulting GW spectrum carries indirect imprints of the heavy sector and the post-inflationary dynamics. This offers an observational window into otherwise inaccessible supermassive particles and provides a powerful probe of high-scale physics beyond the SM.
\end{abstract}

\maketitle

 \preprint{OU-HET 1281} 
	
\section{I\MakeLowercase{ntroduction}}	\label{sec:intro}
Inflation, a phase of accelerated expansion in the early universe, elegantly solves the flatness and horizon problems of standard cosmology \cite{Guth:1980zm, Kazanas:1980tx, Sato:1981qmu, Linde:1981mu, Linde:1983gd, Olive:1989nu, Lyth:1998xn, Riotto:2002yw, Kinney:2003xf, Baumann:2009ds}. This occurs if a scalar field (the inflaton) temporarily dominates the energy density, causing spatial volume to expand quasi-exponentially. The energy transfer mechanism from the inflaton to Standard Model (SM) particles after inflation has important cosmological implications. In the standard picture of reheating, this energy transfer occurs through the coherent oscillation of the inflaton at the end of inflation. The completion of this process marks the beginning of the radiation-dominated era, setting the initial conditions for Big Bang Nucleosynthesis (BBN) \cite{Ichikawa:2005vw, Kawasaki:2000en, Barbieri:2025moq}. Perturbative reheating has been extensively studied in the literature \cite{Abbott:1982hn, Albrecht:1982mp, Kolb:1990vq, Giudice:2000ex, Kolb:2003ke, Cook:2015vqa, Cai:2015soa}, offering a simple and controlled framework for modeling the transition from cold, inflaton dominated universe to the hot thermal bath of SM particles.

However, in many well-motivated inflationary scenarios, the perturbative decay of the inflaton may be inefficient or entirely absent. In such cases, a non-perturbative, explosive particle production mechanism, can efficiently transfer energy from the inflaton to other fields through parametric resonance or tachyonic instability, known as the preheating \cite{Shtanov:1994ce, Dolgov:1989us, Traschen:1990sw, Kofman:1994rk, Boyanovsky:1995ud, Yoshimura:1995gc, Kofman:1997yn}. Unlike perturbative reheating, which proceeds gradually through the coherent oscillation of the inflaton at its minima, preheating involves the rapid, collective excitation of quantum fluctuations in an expanding background. These resonant effects can lead to the copious production of particles within the first few oscillations. Depending on the couplings and structure of the inflationary potential, preheating can excite a wide range of particles. The mechanism of particle production during preheating can differ significantly depending on the spin of the produced species. In particular, scalar particles can undergo broad or narrow parametric resonance depending on the coupling structure and are typically produced more efficiently due to Bose enhancement \cite{Kofman:1997yn,Greene:1997fu}. On the other hand, fermion production is limited by Pauli exclusion, making the resonance narrow and generally less efficient. Nevertheless, under suitable conditions, such as derivative couplings or non-minimal interactions, fermions can still be produced in sizable amounts \cite{Greene:1998nh, Giudice:1999fb, Peloso:2000hy}. These post-inflationary mechanisms not only govern the stage for the thermal history of the universe but can also leave observable imprints in the form of relic particles \cite{Qutub:2017wnf, Garcia:2021iag, Garcia:2022vwm, Haque:2023zhb, Haque:2024zdq}, altered expansion dynamics \cite{Podolsky:2005bw, DiMarco:2021xzk}, and a stochastic background of gravitational waves (GWs) \cite{Khlebnikov:1997di, Easther:2006gt, Easther:2006vd, Garcia-Bellido:2007nns, Garcia-Bellido:2007fiu, Dufaux:2007pt, Dufaux:2008dn, Figueroa:2011ye, Bethke:2013aba, Bethke:2013vca, Figueroa:2017vfa, Ai:2025fqw}. Among the various phenomena that can arise during preheating, the production of supermassive right-handed neutrinos (RHNs), can lead to interesting phenomenological consequences. 

Right-handed neutrinos (RHNs) are a generic ingredient of seesaw mechanisms for generating light neutrino masses \cite{Minkowski:1977sc,Yanagida:1979as,Yanagida:1979gs,GellMann:1979vob,Mohapatra:1979ia,Schechter:1980gr,Schechter:1981cv}. Independently of their role in neutrino mass generation, RHNs can also participate in the generation of the observed baryon asymmetry via leptogenesis \cite{Fukugita:1986hr, Luty:1992un, Pilaftsis:1997jf, Ma:1998dx, Hambye:2000ui, Hambye:2003ka}, although successful leptogenesis does not uniquely fix the RHN mass scale and can be realized over a wide range of masses and dynamical regimes. Once produced in the early Universe, heavy RHNs can significantly affect the thermalization history through their Yukawa interactions with SM particles, even in scenarios where the inflaton does not decay perturbatively into the visible sector. Moreover, irrespective of whether they directly source the baryon asymmetry, massive RHNs can act as efficient emitters of stochastic GWs through graviton bremsstrahlung during their production and subsequent interactions.  The production of GWs through bremsstrahlung has been shown to be a generic and robust source of high-frequency GWs associated directly with the inflaton decay or scattering \cite{Nakayama:2018ptw, Barman:2023ymn, Xu:2024fjl, Datta:2025wfh}, as well as with the decay of superheavy particles generated through various early universe dynamics such as inflaton deacy \cite{Ghoshal:2022kqp, Kanemura:2023pnv, Datta:2024tne, Konar:2025iuk}, primordial black hole (PBH) evaporation \cite{Choi:2024acs}, gravitational production \cite{Inui:2024wgj} or instant preheating \cite{Hu:2024bha, Hu:2024awd}.
Previous studies associated with the graviton bremsstrahlung of RHNs \cite{Ghoshal:2022kqp, Datta:2024tne} have typically assumed that superheavy RHNs, with masses as large as $M_N \gtrsim 10^{15}\, \mathrm{GeV}$, are produced via perturbative inflaton decays. However, in the single field inflation models, the inflaton mass typically falls in $10^{13}\, \mathrm{GeV}$ ballpark. Hence, the perturbative decay in minimal inflationary models is not a viable possibility. The production of heavy RHNs has also been associated with late-time phase transitions \cite{Watkins:1991zt, Konstandin:2011ds, Falkowski:2012fb, Dasgupta:2022isg, Cataldi:2024pgt, Cataldi:2025nac}. In contrast, here we consider a minimal framework in which the inflaton couples only to RHNs. In this setup, nonperturbative particle production during preheating allows for an efficient generation of RHNs even when their masses exceed the inflaton mass, as also pointed out before in \cite{Giudice:1999fb}, thereby providing a simple and self-consistent origin for superheavy RHNs without invoking additional sectors or phase transitions.

To study this process in a concrete setup, we consider the class of $\alpha-$attractor inflationary models \cite{Kallosh:2013lkr, Kallosh:2013hoa, Kallosh:2013yoa, Kallosh:2013pby, Kallosh:2013maa, Galante:2014ifa}. These models allow a wide range of post-inflationary dynamics depending on the steepness of the potential near its minimum, which directly influences the inflaton's oscillatory behavior after inflation. In particular, the effective equation of state (EoS) during the reheating phase, denoted by $\omega_{\rm re}$, can significantly affect both the efficiency of RHN production and the resulting GW spectrum. The value of $\omega_{\rm re}$ determines how rapidly the inflaton energy density redshifts, thereby influencing the onset and efficiency of thermalization. By approximating the reheating phase with a constant $\omega_{\rm re}$, one can derive semi-analytic relations between the duration of reheating and the scalar spectral index $n_s$, connecting post-inflationary physics to Cosmic Microwave Background (CMB) observables \cite{Martin:2010kz, Adshead:2010mc, Mielczarek:2010ag, Dai:2014jja, Martin:2014nya, Drewes:2017fmn, Drewes:2022nhu}. Of particular interest are scenarios with $\omega_{\rm re}\geq 1/3$, where the inflaton energy density redshifts faster than radiation. As we will show, such scenarios are not only a theoretical outcome within steep $\alpha-$attractor potentials but are also favored by current cosmological constraints on $n_s$. In this work, we explore the possibility of probing the presence of such supermassive RHNs through their imprints on the stochastic GW spectrum. This provides an indirect yet promising avenue to test scenarios where the inflaton preferentially couples to heavy states beyond the SM, even in the absence of a direct thermal connection to the visible sector. 

The rest of the article is organized as follows. In section \ref{sec:preheating}, we will briefly discuss the preheating production of super-massive RHNs. In section \ref{sec:obs}, we will discuss the impact of the faster red-shifting of inflaton field and connection of the CMB observables. We discuss the production of matter-antimatter asymmetry through non-thermal leptogenesis in section \ref{sec:YBL}. The GW spectrum produced from the super-massive RHNs will be discussed in section \ref{sec:GWspec} and we have concluded our results in section \ref{sec:cnlsn}.
\section{P\MakeLowercase{roduction of super heavy Right Handed Neutrinos at preheating}}	\label{sec:preheating}
\begin{figure}[h]
    \centering
    \includegraphics[scale=0.45]{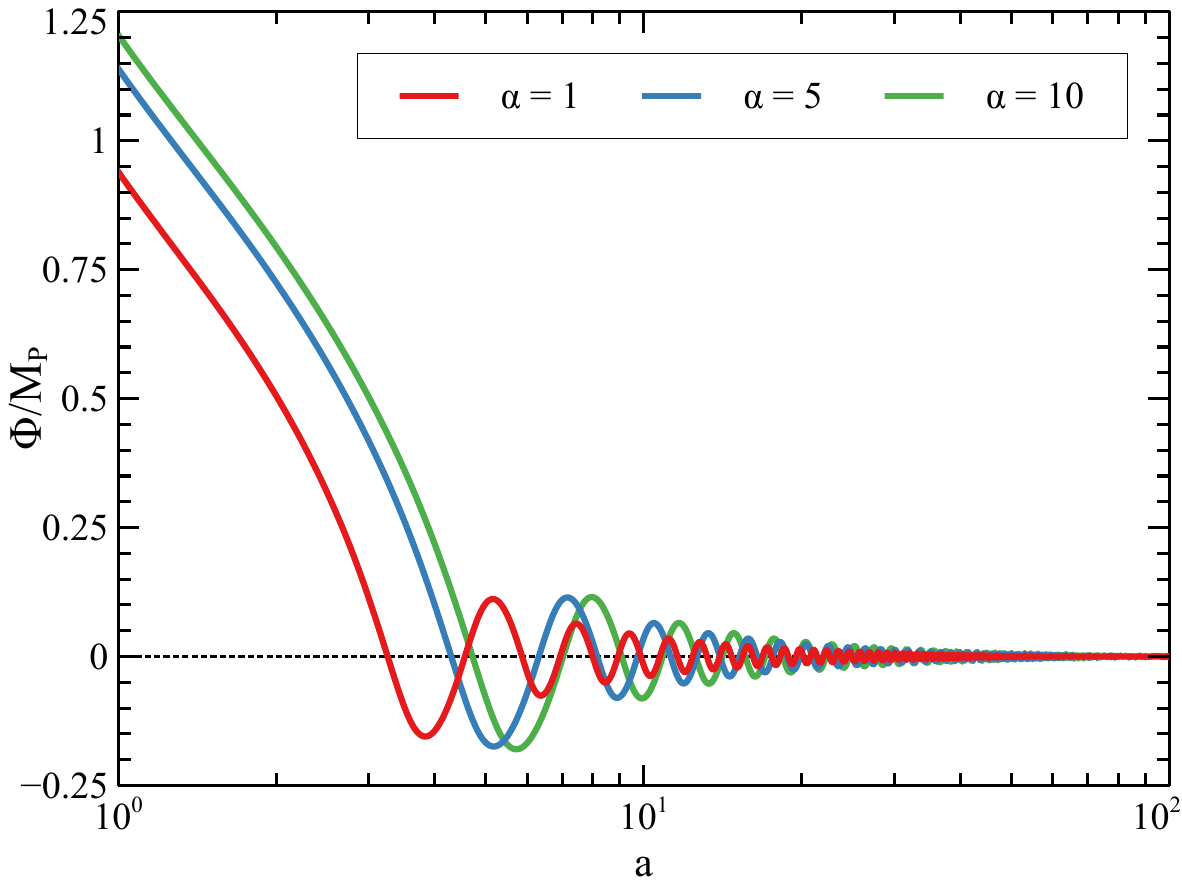}
    \caption{Inflation field value normalised by the reduced Planck mass ($M_P$) as a function of scale factor. }
    \label{fig:fieldvalue}
\end{figure}

In this section, we describe the basic formalism of heavy fermion production during the preheating stage by focusing on the $\alpha-$attractor potential for inflation. A general form of $\alpha-$attractor potential can be written as \cite{Kallosh:2013lkr, Kallosh:2013hoa, Kallosh:2013yoa, Kallosh:2013pby, Kallosh:2013maa, Galante:2014ifa} 
\begin{eqnarray}
    V\left(\phi\right) = \Lambda^4\left[1-e^{-\sqrt{\frac{2}{3\alpha}} \frac{\phi}{M_{P}}}\right]^{2n},
    \label{eq:InfPot}
\end{eqnarray}
where $\phi$ behaves as inflaton, $M_P = 2.44 \times 10^{18} \text{ GeV}$ is the reduced Planck mass, and $\Lambda$ is the mass scale that determines the energy scale of the inflation. $\alpha$ and $n$ are the two independent parameters that determine how steeply the inflaton field rolls and the
shape of the potential near its minimum respectively. $\alpha=1$ and $n=1$ mimics the standard Higgs-Starobinsky inflation potential \cite{Bezrukov:2007ep}. The inflationary stage or {\it slow roll stage}, during which the inflaton potential energy dominates the total energy budget of the universe, is charecterised by the {\it slow-roll parameters} $\epsilon, \eta << 1$ and an exponential growth of the scale factor. For $\alpha-$attractor potential, the end of the inflation is charecterised by $\epsilon = 1$ that corresponds to the field value at the end of the inflation as 
\begin{eqnarray}
   \phi_{\rm end} = \sqrt{\frac{3 \alpha}{2}} M_{P} \ln{\left(\frac{2 n}{\sqrt{3 \alpha}} +1\right)}.
\end{eqnarray}
The equation of motion for inflaton field can be written as 
\begin{eqnarray}
    \frac{d^2 \phi}{dt^2} + 3 H \frac{d\phi}{dt} + \frac{\partial V(\phi)}{\partial \phi} = 0, 
    \label{eq:eom_phi}
\end{eqnarray}
where H is the Hubble parameter. At the end of the inflation, inflaton field oscillates around its minimum. The effective mass of the inflaton is defined as $m_{\phi}^2 = \partial^2_{\phi} V\left(\phi \right)$. In figure~\ref{fig:fieldvalue}, we have shown the solution of the equation of motion of the inflaton for different values of the parameter $\alpha$ where one can note that the field value $\phi_{\rm end}$ at the end of the inflation can be $\sim$ $M_P$ or more for larger $\alpha$. 
Let us assume that the inflaton field interacts only with the heavy RHNs $N_{R}$ with a bare mass $M_N$ via the Yukawa coupling $y_{N} \phi \overline{N^C} N$ and the total mass of the RHN is given by 
\begin{eqnarray}
    m_N(t) = M_{N} + y_{N} \phi(t).
\end{eqnarray}
For clarity, we use "$m$" to denote the effective mass of the RHNs, while "$M$" refers to their bare mass term.
 We start by canonically quantizing the action of the massive field $N_R$ in curved space with Friedmann-Lema\^{i}tre-Robertson-Walker (FLRW) metric. In the system of coordinates in which the line element is given by $ds^2 = a^2(\tau)(d\tau^2 - d \Vec{x}^2)$, where $a$ is the scale factor of the expanding universe and $\tau$ is the conformal time defined as $d\tau = dt/a$, the Dirac equation becomes
\begin{eqnarray}
    \left(\frac{i}{2} \gamma^\mu \partial_{\mu} + i \frac{3}{2} \mathcal{H} \gamma^0 - m_{N}\right) N = 0.
    \label{eq:Dirac1}
\end{eqnarray}
Here $\mathcal{H}=(a^\prime/a^2)$ is the Hubble rate where the prime denotes the derivative with respect to the conformal time $\tau$, and the $\gamma$ matrices are defined in the flat space-time. By defining $\mathcal{N}=a^{3/2} N$, equation~\eqref{eq:Dirac1} can be reduced to the more familiar form
\begin{eqnarray}
    \left(i \gamma^\mu \partial_{\mu} - a m_{N}\right) \mathcal{N}=0.
    \label{eq:EOMN}
\end{eqnarray}
Since $a$ is a function of $\tau$, but not of $\Vec{x}$, spatial translations are symmetries of space-time, and we can separate the variables using the decomposition 
{\small \begin{eqnarray}
    \mathcal{N}(x) = \int \frac{d^3 k}{ (2\pi)^{3/2}} e^{i\Vec{k}.\Vec{x}}\sum \left[u_{r}(k, \tau) a_r(k) + v_{r}(k, \tau) b^\dagger_r (-k)\right],
\end{eqnarray}}
where the summation is over spin of RHNs, and $v_{r}(k) = C \bar{u}^T_{r}(-k)$. We impose the canonical anti-commutation relations on the creation and annihilation operators 
\begin{eqnarray}
    \left\{a_r(k), a_s^\dagger (k^\prime) \right\} = \left\{b_r (k), b_{s}^\dagger (k^\prime) \right\} = \delta_{rs}\delta(\Vec{k} - \Vec{k^\prime}),
\end{eqnarray}
which, together with the quantization conditions, determine the normalization of the spinors, 
\begin{eqnarray}
    u^\dagger_{r}(k, \tau) u_s (k,\tau) = v^\dagger_{r} (k, \tau) v_s(k,\tau) = 2\delta_{rs},\nonumber \\ u^\dagger_r(k,\tau) v_s(k, \tau) = 0.
    \label{eq:creaann}
\end{eqnarray}
The above equations are valid at any conformal time, since they are preserved by the evolution. In the representation in which $\gamma^0 = \begin{pmatrix}
\mathds{1} & 0 \\
0 & \mathds{1}
\end{pmatrix}$ and with the definition of $u \equiv \begin{pmatrix}
u_+ \\
u_-
\end{pmatrix}$, the equation of motion~\eqref{eq:EOMN} can be written as a set of uncoupled second order differential equations,
\begin{eqnarray}
    \left[\frac{d^2}{d\tau^2} + \omega^2 \pm i (a^\prime m_N + a m_N^\prime )\right] u_{\pm} (k) = 0,
    \label{eq:upm}
\end{eqnarray}
with $\omega^2 = k^2 + m_{N}^2 a^2$. We can now write the Hamiltonian as 
\begin{eqnarray}
    H(\tau) &=& \frac{1}{a} \int d^3 x \mathcal{N}^\dagger i \partial_0 \mathcal{N}\\
&=& \frac{1}{a} \int d^3 k \sum_{r} \left\{E_k(\tau) \left[a^\dagger_r(k) a_r(k) - b_r(k)b_r^\dagger(k)\right] + \right. \nonumber  \\
&& \left. F_k(\tau) b_r(-k) a_r(k) + F_k^* (\tau) a^\dagger_{r}(k) b_r^\dagger (-k)\right\}.
\label{eq:Ham}
\end{eqnarray}
By using the equations of motion, we find 
\begin{eqnarray}
    E_{k} &=& k \text{ Re}(u_+^* u_-) - a m_{N} (1-u_+^* u_+), \nonumber \\
    F_{k} &=& \frac{k}{2} (u^2_+ - u^2_-) - a m_N u_+ u_-, \nonumber \\
    E_k^2 + |F_k|^2 &=& \omega^2. 
\end{eqnarray}
Here, we have chosen the momentum $k$ along the third axis, and selected the gamma-matric representation in which $\gamma^3=\begin{pmatrix}
    0 & \mathds{1} \\
    - \mathds{1} & 0
\end{pmatrix}.$ In order to give a ``quasi-particle'' interpretation, we diagonalize the hamiltonian in eq. \eqref{eq:Ham} with a time-dependent {\it Blogolyubov} canonical transformation, and define the new creation and annihilation operators as
\begin{eqnarray}
    \hat{a}(k, \tau) &=& \alpha(k, \tau) a(k) + \beta(k, \tau) b^\dagger (-k), \nonumber \\
    \hat{b}(k, \tau) &=& - \beta^*(k,\tau) a(k) + \alpha^*(k,\tau) b^\dagger (-k).
\end{eqnarray}
Imposing canonical anti-commutation relations on the operators $\hat{a}$ and $\hat{b}$, we find $|\alpha|^2 + |\beta|^2 =1$. For
\begin{equation}
    \frac{\alpha}{\beta} = \frac{E_k + \omega}{F^*_k}, \,\,\,\,\, |\beta|^2 = \frac{|F_k|^2}{2\omega (\omega + E_k)} = \frac{\omega -E_k}{2\omega},
\end{equation}
the normal-ordered hamiltonian in terms of the ``quasi-particle'' operators is diagonal, 
\begin{eqnarray}
    H(\tau) = \frac{1}{a} \int d^3k \sum_{r} \omega(\tau) \left[\hat{a}^\dagger_r(k) \hat{a}_r(k) + \hat{b}^\dagger_r (k) \hat{b}_r(k) \right]. 
\end{eqnarray}
Next, we define a ``quasi-particle'' vacuum, such that $\hat{a} |0_\tau\rangle = \hat{b}|0_\tau\rangle = 0$. Similarly, the initial vacuum $|0\rangle$ is defined as $a|0\rangle = b|0\rangle = 0$. The total number density of produced RHNs up to time $\tau$ is given by the vacuum expectation value of the particle number operator $N$ divided by the physical volume,
\begin{eqnarray}
    n(\tau) = \langle 0| \frac{N}{V}|0\rangle = \frac{1}{2\pi^2 a(\tau)^3} \int _0^\infty dk k^2 |\beta|^2.
\end{eqnarray}
The density of produced particles is then computed by integrating the equations of motion \eqref{eq:upm} with an initial condition at time $\tau_i$ given by 
\begin{eqnarray}
    u_{\pm}(\tau_i) &=& \sqrt{1\pm \frac{m_N a}{\omega}},\\ u_{\pm}^\prime (\tau_i) &=& -i k u_{\mp}(\tau_i) \mp i a m_{N} u_{\pm} (\tau_i).
\end{eqnarray}
This boundary condition corresponds to $E_k = \omega$, $F_k = 0$ at $\tau = \tau_i$ or, in other words to an initial vanishing particle density. 
\begin{figure}[h!]
    \centering
    \includegraphics[scale=0.36]{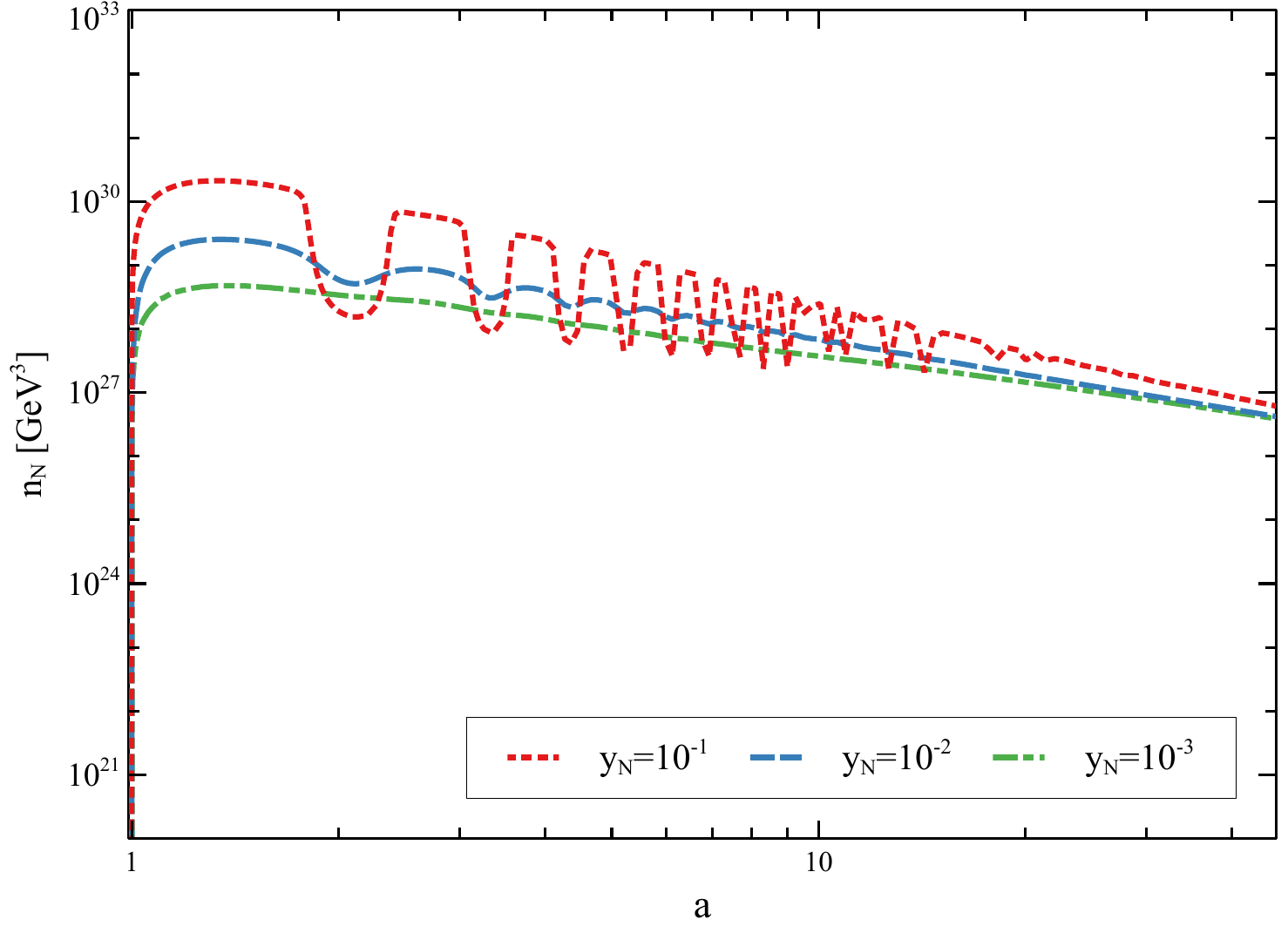}
    \caption{Number density of heavy RHNs during preheating as a function of the scale factor.}
    \label{fig:num_density1}
\end{figure}
In figure~\ref{fig:num_density1}, we have shown the evolution of the number density of heavy RHNs produced during preheating (the first few oscillations) as a function of scale factor for three different benchmark values of the Yukawa coupling $y_N$. One can notice that the production of RHNs reduces with the reduction of $y_N$. Here, we assume the bare mass of the RHN is larger than the mass of the inflaton, {\it i.e.} $M_N/m_{\phi} = 100$.{Here, we have assumed that all three RHNs are heavier than the inflaton. As a result, the perturbative decay of the inflaton into RHNs is kinematically forbidden. On the other hand, once the particle production through preheating stops, the RHNs subsequently decay into SM particles, thereby reheating the universe. Importantly, the resulting reheating temperature remains below than the RHN masses, ensuring that the RHNs stay out-of-equilibrium and satisfy one necessary condition for leptogenesis. During this phase, the inflaton field effectively behaves as a spectator and continues to contribute to the total energy density of the Universe before eventually redshifting away. These aspects are discussed in more detail in later sections.}\\

{Before discussing the implications of these superheavy RHNs, we also want to mention that the backreaction becomes relevant when the energy density of produced fermions approaches that of the inflaton, which happens for sufficiently large values of $q$. Numerical studies show that for $q \leq 10^8$, back reaction has only a minor impact: the particle spectrum changes mainly at low momenta, and the total number density shifts by only a few percent. Even up to $q\sim10^{10}$, back reaction modifies the RHN energy fraction by a factor of two \cite{Giudice:1999fb}. Therefore, throughout the parameter space of interest, neglecting back reaction does not change the number density of RHNs by orders of magnitude, and the main results remain unchanged.}
\section{C\MakeLowercase{onnection to} CMB-\MakeLowercase{observables}}
\label{sec:obs}
By considering the expansion history of the universe between the time the observable CMB modes exited the Hubble radius during inflation and the time they later re-entered, one can extract information about the reheating temperature. The horizon exit of a given mode $k$ can be defined as $k = a_k  H_k$ or it can also be written as 
{\small \begin{eqnarray}
     \ln\left(\frac{a_{\rm end}}{a_k}\right) + \ln\left(\frac{a_{\rm re}}{a_{\rm end}}\right) + \ln\left( \frac{a_{0}}{a_{\rm re}}\right) + \ln \left( \frac{k}{a_0 H_k}\right) = 0,
    \label{eq:Nknre}
\end{eqnarray}}
where $a_{\rm end},\, a_{\rm re},\,\text{and } a_0$ represent the scale factors at the end of the inflation, at the beginning of the radiation domination, and the present day respectively. The above equation can be interpreted in terms of the number of {\it e-folds} $N_k$, the e-folding number from the horizon exit of $k$-mode to the end of inflation, and $N_{\rm re}$, the e-folding number from the end of inflation to the end of the reheating, as follows,
\begin{eqnarray}
    N_k + N_{\rm re} + \ln\left( \frac{a_{0}}{a_{\rm re}}\right) + \ln \left( \frac{k}{a_0 H_k}\right) = 0.
    \label{eq:akHk}
\end{eqnarray}
The third term in Eq.\, \eqref{eq:akHk} can be calculated from the entropy conservation principle between the end of reheating and the present epoch as 
\begin{eqnarray}
    \ln\left(\frac{a_0}{a_{\rm re}}\right) &=& \frac{1}{3}\ln\left(\frac{11 g_{*s}}{43}\right) + \frac{1}{4}\ln \left(\frac{30 \rho_{\rm end}}{g_{*} \pi^2 T_0^4}\right)  \nonumber \\
&& - \frac{3}{4}(1+\overline{\omega}_{re}) N_{re},
\label{eq:entropy}
\end{eqnarray}
where $\rho_{\rm end}$ is the energy density of the inflaton at the end of the inflation, and $T_0$ is the present day temperature of the universe. In deriving the last equation, we used the definition of $N_{\rm re}$ as 
\begin{eqnarray}
    N_{\rm re} = \ln \left(\frac{a_{\rm re}}{a_{\rm end}} \right) = -\frac{1}{3(1+\overline{\omega}_{\rm re})} \ln\left(\frac{\rho_{\rm re}}{\rho_{\rm end}}\right).
    \label{eq:Nre}
\end{eqnarray}
$g_{*s}$ and $g_*$ are the effective relativistic degrees of freedom contributing to entropy density and energy density respectively. The average EoS can be parameterized as $\overline{\omega}_{\rm re} = \frac{n-1}{n+1}$ \cite{Lozanov:2016hid}.  In this context, the most important CMB observables that are often used to constrain the inflationary model parameters are the amplitude of the scalar perturbations $A_s$, the tensor-to-scalar ratio $r$, and the spectral index $n_s$. These observables are evaluated at some reference scale known as the {\it pivot scale}, i.e., a specific mode of the inflaton fluctuation with a comoving wave number $k$. By calculating the {\it slow-roll parameters} and the value of $H$ when the mode $k$ crosses the horizon, these observables can be written as 
\begin{eqnarray}
n_s = 1-6\epsilon_k + 2 \eta_k,\,\, r = 16\epsilon_k,\,\,
A_s = \frac{2 H_k^2}{ \pi^2 M_P^2 r}.
\label{eq:CMBobs}
\end{eqnarray}
By using the definition of $H_k$ and $\rho_{\rm end}$ given in Eq.\,\eqref{eq:CMBobs} and Eq.\,\eqref{eq:Nre} respectively, Eq. \eqref{eq:akHk} can now be rewritten as 
\begin{eqnarray}
 N_{\rm re} &=&  \frac{4}{3 \overline{\omega}_{re}-1}\left[N_{k}  + \frac{1}{3}\ln\left(\frac{11 g_{*s}}{43}\right) + \frac{1}{4}\ln \left(\frac{40 }{g_{*} \pi^2}\right)\right.  \nonumber \\ && \left.+ \ln \left(\frac{k}{a_0 T_0}\right) + \frac{1}{2}\ln\left(\frac{2 \sqrt{V_{\rm end}}}{\pi^2 M_P^2 r A_s}\right)\right],
 \label{eq:NreRHS}
\end{eqnarray}
where we have used the energy density at the end of inflation $\rho_{\rm end} = (4/3) V_{\rm end}$. During our numerical analysis, we adopt the pivot scale value \( k/a_0 = 0.05\, \text{Mpc}^{-1} \). According to Planck data~\cite{Planck:2018jri}, the observed scalar amplitude is \( A_s = 2.099 \times 10^{-9} \), and the present temperature of the universe is \( T_0 = 2.7\, \text{K} \).
\begin{figure*}[hbt!]
    \centering
    \includegraphics[scale=0.24]{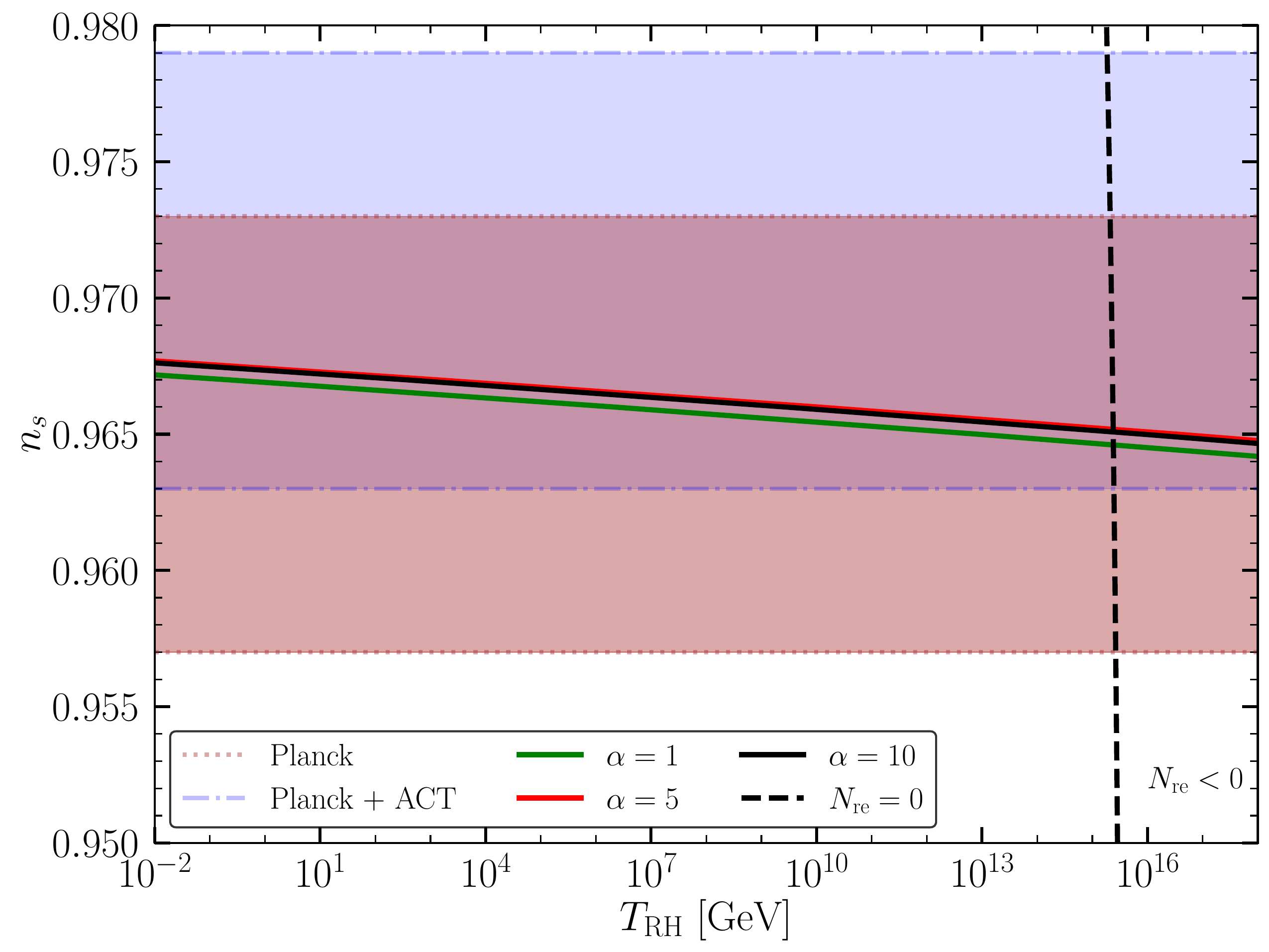}
    \caption{Prediction of CMB spectral index ($n_s$) as a function of the reheating temperature ($T_{\rm RH}$) for different values of $\alpha$ and $n=3$.}
    \label{fig:nsTRH}
\end{figure*}
It is important to emphasize that the right-hand side of Eq.\,\eqref{eq:NreRHS} depends only on the inflationary model parameters and CMB observables. In contrast, the left-hand side depends not only on the model parameters but also on the coupling between the inflaton and the radiation bath, as defined in Eq.\,\eqref{eq:Nre}. Given a specific interaction between the inflaton and bath particles, along with inflationary parameters \( \alpha \) and \( n \), one can straightforwardly compute the reheating energy density \( \rho_{\rm re} \), and thereby determine the reheating temperature \( T_{\rm RH} \). Equating Eq.\,\eqref{eq:Nre} with Eq.\,\eqref{eq:NreRHS} then enables us to extract the corresponding predictions for the observables \( n_s \) and \( r \). {However, these predictions depend on the average EoS, $\omega_{\rm re}$, during the reheating phase. Consequently, different values of the parameter $n$ lead to different correlations between the scalar spectral index $n_s$ and the reheating temperature $T_{\rm RH}$. For $n=1$, the EoS corresponds to a matter-like behavior with $\omega_{\rm re}=0$, and $n_s$ increases with $T_{\rm RH}$. This behavior changes for $n>1$, as also discussed in Ref.~\cite{Cook:2015vqa, Antoniadis:2025pfa, Biswas:2025adi}. In the present framework, since we assume that the inflaton couples only to heavy RHNs and its perturbative decay is kinematically forbidden, the inflaton energy density must redshift faster than that of radiation; otherwise, the Universe would remain dominated by the inflaton. Such scenarios can be realized for $n \geq 2$. The evolution of the different energy densities is illustrated in the bottom panel of Fig.~\ref{fig:story}.} In figure~\ref{fig:nsTRH},  we have shown the variation of $n_s$ as a function of $T_{RH}$ for three different benchmark values of $\alpha$ and fixed $n=3$. The red shaded region is the allowed range of $n_s$ from Planck data whereas the blue shaded region is the allowed range from the recently released Planck+ACT data \cite{ACT:2025tim, ACT:2025fju}. The most important take away is that the lower $T_{\rm RH}$ corresponds to the larger values of $n_s$ or in other words the central value of Planck+ACT data favors the lower $T_{\rm RH}$.

\section{P\MakeLowercase{roduction of matter-antimatter asymmetry}}	\label{sec:YBL}
\begin{figure*}[htbp]
  \centering
  \includegraphics[width=13.5cm, height=4cm]{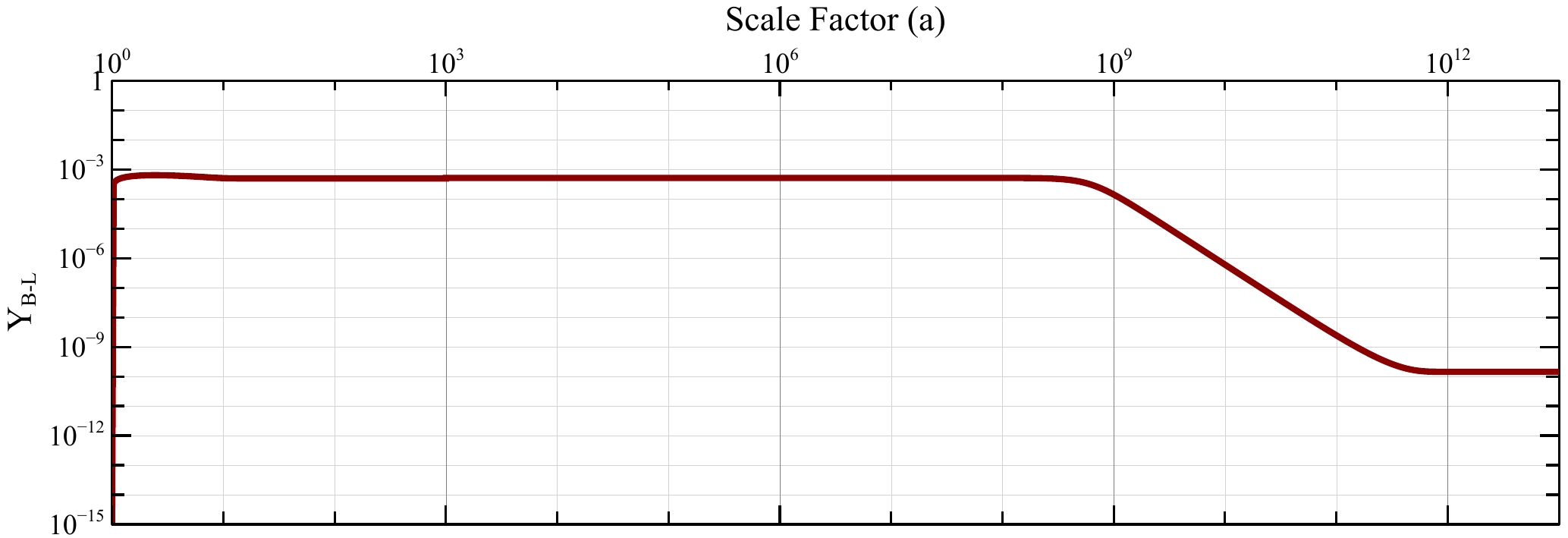}\\
  \includegraphics[width=13.5cm, height=7cm]{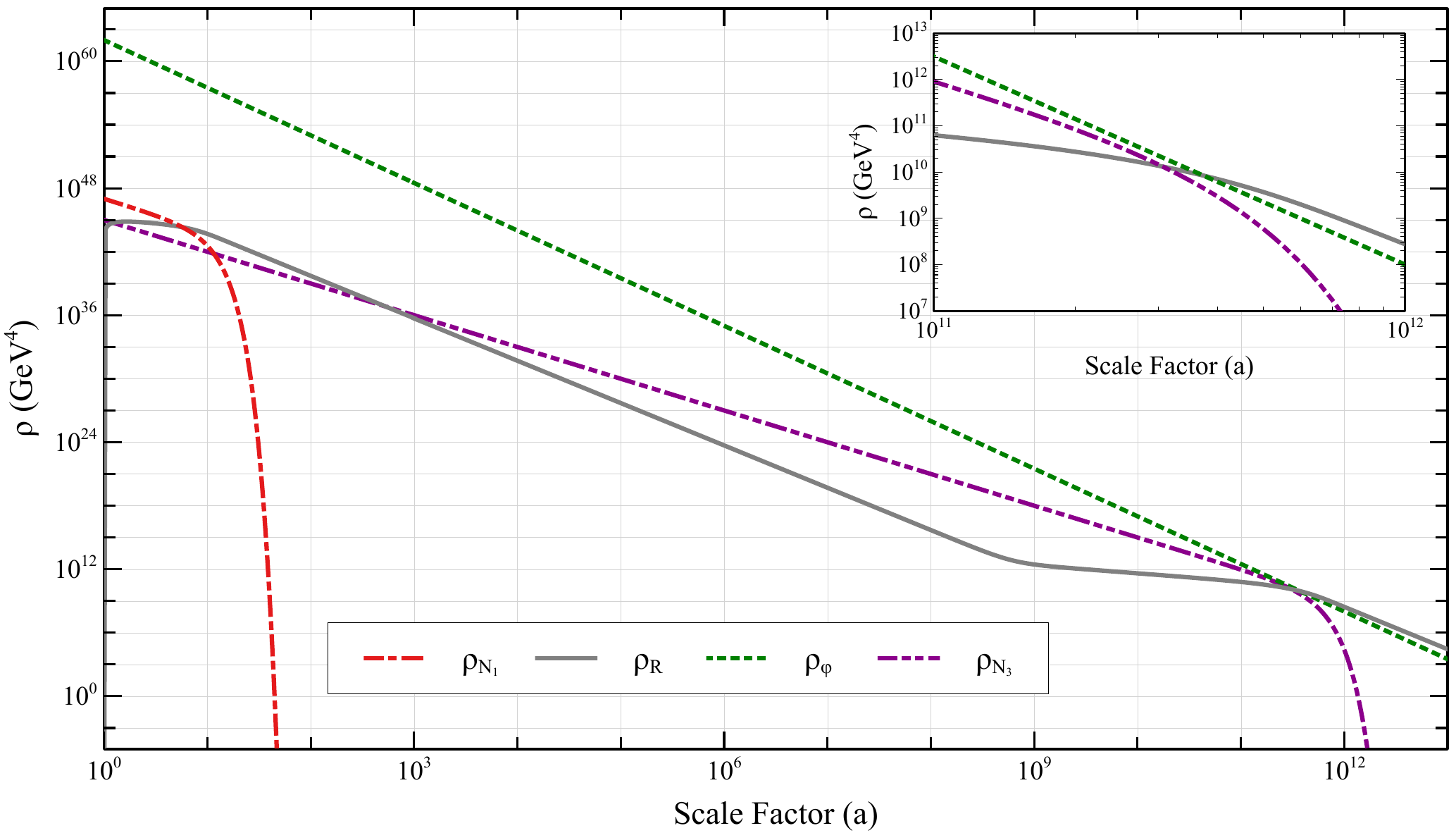}
  \caption{Evolution of energy densities of different species in the early universe for fixed benchmark values of the parameters. The inflationary parameters are taken as $n=3$ and $\alpha=1$, the asymmetry parameter $\varepsilon_{\ell} = 0.1$, the mass of the RHNs are $M_1\approx M_2 = 10^{15} \text{ GeV and } M_3=10^{16} \text{ GeV}$, and the decay widths of the RHNs are $\Gamma_{N_{1,2}}= 10^{10} \text{ GeV}$ and $\Gamma_{N_3} = 10^{-12} \text{ GeV}$.}
  \label{fig:story}
\end{figure*}

Before evaluating the GW spectrum resulting from the long-lived RHN decay, we briefly outline the mechanism of non-thermal leptogenesis in this framework. Once produced via preheating, the heavy RHNs would decay into the SM leptons and Higgs bosons through the Yukawa interactions describe by  the Lagrangian: 
\begin{align}
	-\mathcal{L_{\rm N}}= \overline{\ell}_{L_\alpha} (Y_{\nu})_{\alpha i} \tilde{H} N_{i}+ \frac{1}{2}  \overline{N_{i}^c}(M_{N}^{ij}) N_j+ h.c.,
	\label{seesaw}
\end{align}
{It is important to note that the framework considered here contains three RHNs, and we assume a hierarchical mass spectrum given by $M_{N_{1,2}} \ll M_{N_3}$. We further take the Yukawa couplings associated with $N_3$ to be significantly smaller than those of $N_{1,2}$, such that $N_3$ is effectively decoupled from light neutrino mass generation. In this setup, the observed neutrino oscillation data can be accommodated with only two heavy RHNs. As a result, the lightest active neutrino becomes almost massless. The interactions of $N_1$ and $N_2$ can then give rise to CP-violating two-body decays into (anti-)lepton and (anti-)Higgs pairs. While the long-lived nature of $N_3$ implies that it does not directly participate in either light neutrino mass generation or the primary production of the lepton asymmetry, as discussed later, it nevertheless plays an important role in the generation of GWs and in the dilution of the asymmetry through entropy injection during its decay.}
As discussed earlier, these supermassive RHNs never thermalize with the bath particles, and the out-of-equilibrium decay of the $N_i$ generates a finite amount of CP asymmetry, parameterized by 
		\begin{equation}
		\varepsilon_\ell^i= \frac{\Gamma(N_i\to \ell_L+H)-\Gamma(N_i\to \bar{\ell}_L+H^\dagger)}{\Gamma(N_i\to \ell_L+H)+\Gamma(N_i\to \bar{\ell}_L+H^\dagger)},
	\end{equation}
	where the denominator denotes the total decay width of the RHN $N_i$ and is given by (at tree level): 
	\begin{align}
		\Gamma_{N_i}= 
		M_{N_i} \frac{ (Y_{\nu}^\dagger Y_{\nu})_{ii}}{8 \pi}.
        \label{eq:deacyN}
	\end{align}
The resulting lepton asymmetry is partially converted into a baryon asymmetry through electroweak sphaleron transitions, which are active before the electroweak phase transition. In order to compute the amount of $B-L$ asymmetry produced from the CP-violating, out-of-equilibrium decay of the lightest RHN, as well as to track evolutions of all the other components over time, it is necessary to solve the coupled Boltzmann equations (BE) governing the number densities and the $B-L$ asymmetry. They are given by 
\begin{align}
    &\dot{\rho_{\phi}} + 3 \mathcal{H} (1+\overline{\omega}_{re}) \rho_\phi =  0,
    \label{eq:BErhophi}
    \\
   &\dot{\rho}_{N_{1,2}} + 3 \mathcal{H} \rho_{N_{1,2}} =   - \Gamma_{N_{1,2}} \rho_{N_{1,2}}, 
    \label{eq:BErhoN1R}
    \\
     &\dot{\rho}_{N_3} + 3 \mathcal{H} \rho_{N_3} =  - \Gamma_{N_3} \rho_{N_3}, 
    \label{eq:BErhoN2R}
    \\
    &\dot{\rho}_{R} + 4 \mathcal{H} \rho_R =  \Gamma_{N_{1,2}} \rho_{N_{1,2}} + \Gamma_{N_3} \rho_{N_3}, 
    \label{eq:BErhoRad}
    \\
    &\dot{n}_{B-L} + 3 \mathcal{H} n_{B-L} = - \frac{\varepsilon_{\ell} \rho_{N_1} \Gamma_{N_1}}{M_{N_1}},
    \label{eq:nBL}
\end{align}
where $\Gamma_{N_i}$ is the the decay rate of $N_i$ via neutrino-Yukawa interaction as shown in equation~\eqref{eq:deacyN} and $\mathcal{H}$ corresponds to the Hubble expansion rate which can be defined as $\mathcal{H}=\sqrt{{\rho_{\rm tot}}/{3 M_{P}^2}}$, $\rho_{\rm tot}$ is the total energy density of the universe. Note that the evolution equation for the inflaton field ($\phi$) does not include any decay term. In other words, at the end of inflation, the inflaton behaves as a spectator field. The production of the thermal bath is instead driven by the decay of heavy RHNs generated during the preheating stage. In the lower panel of figure~\ref{fig:story}, we show the evolution of the energy densities of the inflaton, RHNs, and radiation as functions of the scale factor, while the upper panel illustrates the evolution of the $B-L$ asymmetry. It may be noted that such a scenario requires a substantial hierarchy in the lifetimes of the RHNs. {It is also important to point out that the asymmetry was mainly generated by the decay of the lightest RHN $N_1$. The decay of $N_3$ doesn't contribute to the asymmetry generation. However, its late-time entropy injection into the SM plasma dilutes the total asymmetry as shown in the upper panel of figure~\ref{fig:story}.} During the decay of $N_3$, the radiation energy density decreases more slowly than in the standard case where $\rho_R \propto a^{-4}$. This is due to the non-trivial redshift behavior of Hubble parameter $\mathcal{H} \propto a^{-9/4}$ during the $\phi$-dominated phase, resulting in $\rho_R \propto a^{-3/4}$. This effect is reflected in the plateau-like behavior of $\rho_R$ in the plot. In the next section, we will discuss the generation of GWs via bremsstrahlung from this same decay process.

Before going into the details of GW spectrum, we note that if the heaviest RHN is stable, it can serve as a viable superheavy dark matter candidate. In such a case, its production during preheating could naturally yield the correct relic abundance, with significant implications for the timing of matter-radiation equality. Interestingly, the dark matter abundance can become comparable to the radiation energy density near the observed equality temperature without requiring significant fine-tuning. A detailed exploration of this possibility and its observational consequences is left for future work.
\section{G\MakeLowercase{ravitational Wave Spectrum}}	\label{sec:GWspec}
\begin{figure}[htbp]
    \centering
    \includegraphics[scale=0.12]{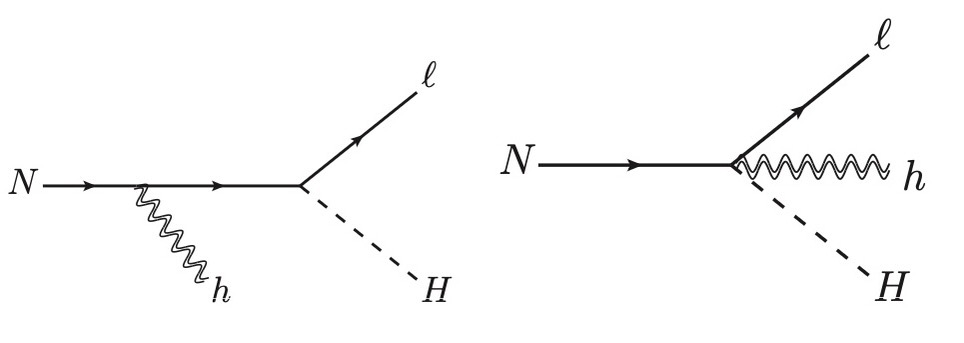}\,\includegraphics[scale=0.12]{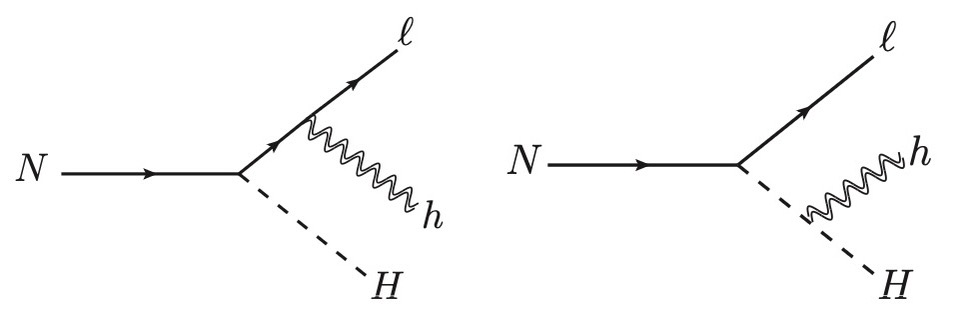}
    \caption{Feynman diagram representing the bremsstrahlung production of graviton.}
    \label{fig:fnmn}
\end{figure}

\begin{figure}[htbp]
    \centering
    \includegraphics[scale=0.43]{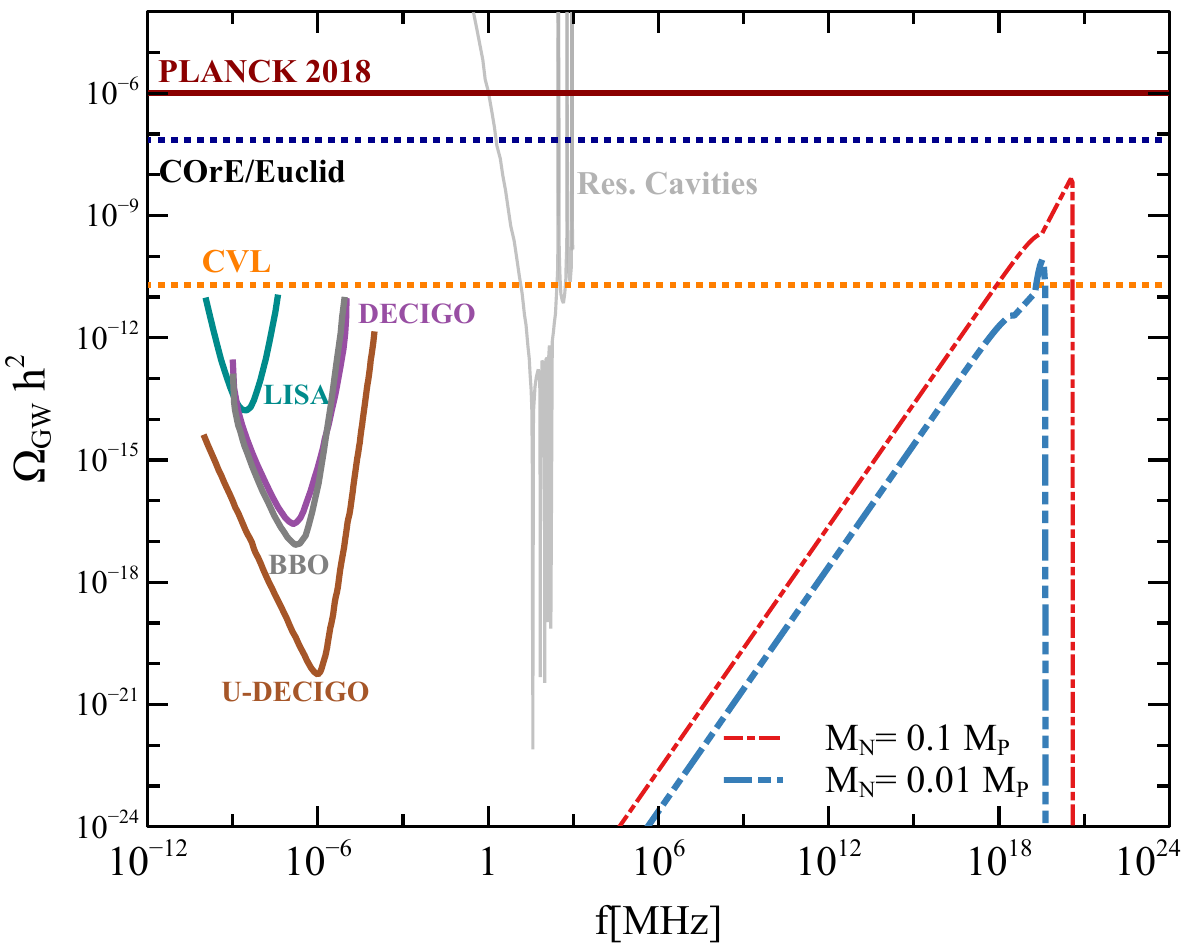}  
    \caption{ GW spectrum arising from graviton bremsstrahlung during super massive RHN $(N_3)$ decaying to lepton doublet and Higgs. We show the GW spectrum for $M=0.1\, M_P \text{ Gev}$ (red) and $M=0.01\, M_P \text{ Gev}$ (cyan).}
    \label{fig:GW}
\end{figure}

In this section, we present the resulting GW spectrum generated through bremsstrahlung emission of gravitons during the decay of the heaviest right-handed neutrino (RHN), $N_3$. Unlike preheating-induced GWs sourced by inflaton dynamics or field inhomogeneities, this mechanism arises from quantum radiation emitted by individual, non-relativistic RHNs as they decay into higgs and leptons as shown in the Feynman diagrams in figure~\ref{fig:fnmn}. Due to the large mass and non-thermal origin of $N_3$, the emitted GWs populate the high-frequency regime. To begin with, the evolution equation of the graviton energy density can be expressed as 
\begin{align}
    \dot{\rho}_{\rm GW} + 4 \mathcal{H} \rho_{\rm GW} = \int \frac{d^3 \kappa}{(2\pi)^3} 2 \kappa \mathcal{F}_{\rm GW} (\kappa,t),
    \label{eq:rhoGW}
\end{align}
where $\mathcal{F}_{\rm GW}$ depends on the production rate of gravitons from RHNs and can be expressed as 
\begin{eqnarray}
    \mathcal{F}_{\rm GW} &=& \frac{1}{4\kappa} \int \Pi_{i=1}^3 \frac{d^3 p_i}{(2\pi)^3 2E_{p_i}}  |\mathcal{M}_{\rm GW}|^2 f_{\rm N_3}(p_1) \times \nonumber \\
    && (2\pi)^4 \delta(P_1-P_2-P_3-K),
\end{eqnarray}
where the four momenta of RHN, $\ell$, $H$, and graviton are $P_1,\, P_2,\, P_3,\, \text{and } K$ respectively whereas the magnitudes of the three momentum vectors are $p_1,\,p_2,\,p_3,\,\text{and } \kappa$. $f_{N_3}$ is the distribution function of the heaviest RHN. The matrix element squared for the given process can be expressed as \cite{Murayama:2025thw},
 \begin{eqnarray}
    |\mathcal{M}_{\rm GW}|^2 &=& \frac{16 |Y_\nu|^2}{ 2 x_{g}^2} \left(\frac{\kappa}{8}\right)^2 M_{N_3}^2 \times \nonumber \\
    && (1-x_{g})(2-x_{g} - x_{g} x_\ell),
    \label{eq:matrx}
\end{eqnarray}
with $x_{g} = \frac{\kappa M_{N_3}}{2}$ and $x_{\ell} = \frac{p_2 M_{N_3}}{2}$. By integrating over the phase space, the equation~\eqref{eq:rhoGW} can be written as 
\begin{eqnarray}
    \dot{\rho}_{\rm GW} + 4 \mathcal{H} \rho_{\rm GW} = \rho_{N_3} \Gamma_{N_3}^{\rm GW},
    \label{eq:rhoGW2}
\end{eqnarray}
where energy density of the $\rho_{N_3}$ is derived from the distribution function as $\rho_{N_3} = \int \frac{d^3 p_1}{(2\pi)^3} E_{p_1} f_{N_3} (p_1)$ and $\Gamma_{N}^{\rm GW}$ is the three body decay width defined as 
\begin{eqnarray}
     \Gamma_{N}^{\rm GW} &=& \frac{1}{(2\pi)^5} \frac{1}{2 M_{N_3}} \int \frac{d^3 p_2}{2E_{p_2}} \frac{d^3 p_3}{2E_{p_3}}\nonumber \\
     && \frac{|\mathcal{M}_{\rm GW}|^2}{E_{\kappa}}\times \delta(M_{N_3} - E_{p_2} - E_{p_3} -E_{\kappa}).
     \label{eq:N3GWdecay}
\end{eqnarray}
 We have integrated the equation~\eqref{eq:rhoGW} and obtain the GW spectrum as $\Omega_{GW} h^2 = \frac{h^2}{\rho_c^0}\frac{d\rho_{GW}^0}{d\ln \kappa^0}$ where $\rho_c^0$ and $\rho_{\rm GW}^0$ are the critical energy density and gravitational energy density today respectively, $\kappa^0 = 2\pi f_{GW}$ is the GW wavenumber today, $f_{GW}$ the GW frequency today.  Due to the super heavy nature of $M_3$, here, we have ignored the mass of SM particles. The calculation of lepton asymmetry is highly sensitive to the decay of \(N_3\), as it is expected to dilute the produced asymmetry in the earlier epoch.
The GW spectrum is expected to depend sensitively on \(M_3\), since the GWs are generated exclusively through single graviton emission (\textit{bremsstrahlung}) from \(N_3\). To gain insight into how the GW spectrum depends on the relevant parameters, we adopt the approximation of instantaneous decay of RHNs at scale factor $a_*$,
\begin{eqnarray}
    \Omega_{\rm GW} h^2 &=& \Omega_{\gamma}^0 h^2 \frac{Y_{\nu}^2}{64 \pi^2} \left(\frac{M_{N_3}}{M_{P}}\right)^2 \frac{\rho_{N_3}^{*}}{\rho_\gamma^0} \left(\frac{a_*}{a_0}\right)^4  \frac{M_{N_3}}{2} \times \nonumber \\ && \frac{1}{\Gamma_{N_3}}\frac{2\kappa_*}{M_{N_3}} \left(\frac{2\kappa_*}{M_{N_3}} - 2\right)^2 \left(1-\frac{2 \kappa_*}{M_{N_3}}\right).
\end{eqnarray}
where $a_0$ is the present scale factor and $\kappa_* = \frac{\kappa_0 a_0}{a_*}$. {The derivation of the above equation is shown in appendix \ref{app}.} One can see the maximum magnitude of the GW spectrum as it depends only on $M_{N_3}$ as, 
{\small \begin{eqnarray}
    \Omega_{\rm GW} h^2\Big|_{\rm peak} = 7.89\times 10^{-11} \left(\frac{M_{N_3}}{0.01 M_{P}}\right)^2.
\end{eqnarray}}
In figure~\ref{fig:GW}, we have shown the resulting GW spectrum as a function of frequency for two different values of the $M_{N_3}$. We have also shown the future predictions from GW detectors such as LISA \cite{Baker:2019nia}, BBO \cite{Corbin:2005ny, Crowder:2005nr, Harry:2006fi}, DECIGO, and Ultimate DECIGO (U-DECIGO) \cite{Seto:2001qf, Kawamura:2020pcg, Kawamura:2006up} along with the current cosmological constraints {from CMB and BBN combined \cite{Yeh:2022heq}} as well as the future projections from resonant cavity techniques~\cite{Berlin:2021txa, Herman:2022fau},  COrE~\cite{COrE:2011bfs}/Euclid~\cite{EUCLID:2011zbd} and Cosmic Variance Limited (CVL) surveys~\cite{Ben-Dayan:2019gll}. It is important to note that the GWs generated via bremsstrahlung from supermassive RHNs predominantly lie in the high-frequency regime, well beyond the sensitivity range of conventional interferometric detectors such as LISA, BBO, or DECIGO. As a result, direct detection of these signals through such experiments remains challenging. However, future cosmological observations, including those from missions like Euclid, COrE, and CVL, are expected to offer significantly improved sensitivity to the effective number of relativistic species, \( N_{\rm eff} \), thereby providing an indirect avenue to probe such high-frequency GW backgrounds and test scenarios like the one presented in this work.
\section{C\MakeLowercase{onclusion and }D\MakeLowercase{iscussions}}
\label{sec:cnlsn}
In this work, we have explored the GW signatures arising from the decay of supermassive RHNs produced during the preheating phase after inflation in the context of $\alpha-$attractor model. Unlike typical GW sources associated with inflaton fragmentation or bubble collisions, our focus was on a quantum mechanical process: bremsstrahlung of gravitons emitted during the decay of RHNs into SM particles. We considered a minimal setup where the inflaton couples exclusively to heavy RHNs, leading to their production via parametric resonance. Our analysis shows that while the $B\!-\!L$ asymmetry generated via standard non-thermal leptogenesis can be significantly affected due to the late-time entropy production by the heaviest RHN decay. However, the resulting GW spectrum from the same RHN depends sensitively on its mass. The resulting GW signals are found to lie in a frequency range that is inaccessible to current and near-future interferometric detectors such as LISA, BBO, or DECIGO. However, they may leave imprints on cosmological observables, such as an excess in the effective number of relativistic species, \( N_{\rm eff} \). Upcoming cosmological probes, including Euclid and CVL surveys, are expected to achieve the sensitivity required to constrain such high-frequency backgrounds indirectly. This opens up the exciting possibility of probing hidden sectors and post-inflationary dynamics through precision cosmology.
\section*{Acknowledgements}
DN would like to thank Sougata Ganguly for the useful discussion. This work is supported by JSPS Grant-in-Aid for JSPS Research Fellows No. 24KF0238. S.K. is also supported, in part, by Grants-in-Aid for Scientific Research (KAKENHI), No. 23K17691 and No. 20H00160.
\appendix
\section{Derivation of the abundance of the GW energy density}
\label{app}

The evolution of the GW energy density $\rho_{\rm GW}$ in an expanding Universe
is governed by the Boltzmann equation
\begin{equation}
\dot{\rho}_{\rm GW} + 4H \rho_{\rm GW}
=
\rho_{N_3}\,\Gamma_{N_3}^{\rm GW},
\label{eq:rhoGW_app}
\end{equation}
where $H$ is the Hubble parameter, $\rho_{N_3}$ denotes the energy density of
the decaying RHN, and $\Gamma_{N_3}^{\rm GW}$ is the partial decay rate for
graviton emission.
The term $4H\rho_{\rm GW}$ accounts for the redshifting of GWs as radiation.

The formal solution of eq.~\eqref{eq:rhoGW_app} evaluated at the present epoch
$t_0$ is
\begin{equation}
\rho_{\rm GW}^0
=
\int_{t_{\rm M}}^{t_{\rm D}} dt\;
\rho_{N_3}(t)\,\Gamma_{N_3}^{\rm GW}
\left(\frac{a(t)}{a_0}\right)^4 ,
\label{eq:rhoGW_int}
\end{equation}
where $t_{\rm M}$ and $t_{\rm D}$ denote the onset and completion of the RHN decay process, respectively, and the factor $(a(t)/a_0)^4$ redshifts the GW energy density from the time of production to today. The term $Y_\nu^2 M_{N_3}^2$ appears through the matrix amplitude square given in eq.~\eqref{eq:matrx}. The present-day GW abundance normalized to the photon energy density can be written as
{\small \begin{eqnarray}
h^2\Omega_{\rm GW}
&=&
h^2\Omega_\gamma^0
\frac{Y_\nu^2}{64\pi^2}
\frac{M_{N_3}^2}{M_P^2}
\int_{t_{\rm M}}^{t_{\rm D}} dt\;
\frac{a^4(t)}{a_0^4}
\frac{\rho_{N_3}(t)}{\rho_\gamma^0}\,\Theta\!\left(1-\frac{2\kappa}{M_{N_3}}\right)
\nonumber\\
&&
\times \kappa
\left(\frac{2\kappa}{M_{N_3}}-2\right)^2
\left(1-\frac{2\kappa}{M_{N_3}}\right).
\label{eq:OmegaGW_decay_app}
\end{eqnarray}}
The prefactor $h^2\Omega_\gamma^0$ arises from normalizing the GW energy density
to the present photon abundance. Since the RHNs are non-relativistic and decay over a time interval much shorter
than the Hubble time, we adopt the instantaneous decay approximation and evaluate
the integral at the decay time $t_* \simeq \Gamma_{N_3}^{-1}$.
This yields
\begin{eqnarray}
\Omega_{\rm GW} h^2
&=&
\Omega_{\gamma}^0 h^2
\frac{Y_{\nu}^2}{64 \pi^2}
\left(\frac{M_{N_3}}{M_P}\right)^2
\frac{\rho_{N_3}^{*}}{\rho_\gamma^0}
\left(\frac{a_*}{a_0}\right)^4
\frac{M_{N_3}}{2}
\frac{1}{\Gamma_{N_3}}
\nonumber\\
&&\times
\frac{2\kappa_*}{M_{N_3}}
\left(\frac{2\kappa_*}{M_{N_3}} - 2\right)^2
\left(1-\frac{2 \kappa_*}{M_{N_3}}\right).
\end{eqnarray}
\bibliographystyle{utphys}
\bibliography{references}

@article{Planck:2018jri,
    author = "Akrami, Y. and others",
    collaboration = "Planck",
    title = "{Planck 2018 results. X. Constraints on inflation}",
    eprint = "1807.06211",
    archivePrefix = "arXiv",
    primaryClass = "astro-ph.CO",
    doi = "10.1051/0004-6361/201833887",
    journal = "Astron. Astrophys.",
    volume = "641",
    pages = "A10",
    year = "2020"
}

@article{Kofman:1997yn,
    author = "Kofman, Lev and Linde, Andrei D. and Starobinsky, Alexei A.",
    title = "{Towards the theory of reheating after inflation}",
    eprint = "hep-ph/9704452",
    archivePrefix = "arXiv",
    reportNumber = "IFA-97-28, SU-ITP-97-18",
    doi = "10.1103/PhysRevD.56.3258",
    journal = "Phys. Rev. D",
    volume = "56",
    pages = "3258--3295",
    year = "1997"
}

@article{Kallosh:2013maa,
    author = "Kallosh, Renata and Linde, Andrei",
    title = "{Non-minimal Inflationary Attractors}",
    eprint = "1307.7938",
    archivePrefix = "arXiv",
    primaryClass = "hep-th",
    doi = "10.1088/1475-7516/2013/10/033",
    journal = "JCAP",
    volume = "10",
    pages = "033",
    year = "2013"
}

@article{Bezrukov:2007ep,
    author = "Bezrukov, Fedor L. and Shaposhnikov, Mikhail",
    title = "{The Standard Model Higgs boson as the inflaton}",
    eprint = "0710.3755",
    archivePrefix = "arXiv",
    primaryClass = "hep-th",
    doi = "10.1016/j.physletb.2007.11.072",
    journal = "Phys. Lett. B",
    volume = "659",
    pages = "703--706",
    year = "2008"
}

@article{Barman:2023ymn,
    author = "Barman, Basabendu and Bernal, Nicol\'as and Xu, Yong and Zapata, \'Oscar",
    title = "{Gravitational wave from graviton Bremsstrahlung during reheating}",
    eprint = "2301.11345",
    archivePrefix = "arXiv",
    primaryClass = "hep-ph",
    doi = "10.1088/1475-7516/2023/05/019",
    journal = "JCAP",
    volume = "05",
    pages = "019",
    year = "2023"
}

@article{Kanemura:2023pnv,
    author = "Kanemura, Shinya and Kaneta, Kunio",
    title = "{Gravitational waves from particle decays during reheating}",
    eprint = "2310.12023",
    archivePrefix = "arXiv",
    primaryClass = "hep-ph",
    reportNumber = "OU--HET--1206, OU-HET-1206",
    doi = "10.1016/j.physletb.2024.138807",
    journal = "Phys. Lett. B",
    volume = "855",
    pages = "138807",
    year = "2024"
}

@article{Minkowski:1977sc,
    author = "Minkowski, Peter",
    title = "{$\mu \to e\gamma$ at a Rate of One Out of $10^{9}$ Muon Decays?}",
    reportNumber = "Print-77-0182 (BERN)",
    doi = "10.1016/0370-2693(77)90435-X",
    journal = "Phys. Lett. B",
    volume = "67",
    pages = "421--428",
    year = "1977"
}

@article{Yanagida:1979as,
    author = "Yanagida, Tsutomu",
    editor = "Sawada, Osamu and Sugamoto, Akio",
    title = "{Horizontal gauge symmetry and masses of neutrinos}",
    reportNumber = "KEK-79-18-95",
    journal = "Conf. Proc. C",
    volume = "7902131",
    pages = "95--99",
    year = "1979"
}

@article{Yanagida:1979gs,
    author = "Yanagida, Tsutomu",
    title = "{Horizontal Symmetry and Mass of the Top Quark}",
    reportNumber = "TU/79/196",
    doi = "10.1103/PhysRevD.20.2986",
    journal = "Phys. Rev. D",
    volume = "20",
    pages = "2986",
    year = "1979"
}

@article{Mohapatra:1979ia,
    author = "Mohapatra, Rabindra N. and Senjanovic, Goran",
    title = "{Neutrino Mass and Spontaneous Parity Nonconservation}",
    reportNumber = "MDDP-TR-80-060, MDDP-PP-80-105, CCNY-HEP-79-10",
    doi = "10.1103/PhysRevLett.44.912",
    journal = "Phys. Rev. Lett.",
    volume = "44",
    pages = "912",
    year = "1980"
}

@article{Schechter:1980gr,
    author = "Schechter, J. and Valle, J. W. F.",
    title = "{Neutrino Masses in SU(2) x U(1) Theories}",
    reportNumber = "SU-4217-167, COO-3533-167",
    doi = "10.1103/PhysRevD.22.2227",
    journal = "Phys. Rev. D",
    volume = "22",
    pages = "2227",
    year = "1980"
}

@article{Schechter:1981cv,
    author = "Schechter, J. and Valle, J. W. F.",
    title = "{Neutrino Decay and Spontaneous Violation of Lepton Number}",
    reportNumber = "SU-4217-203, COO-3533-203",
    doi = "10.1103/PhysRevD.25.774",
    journal = "Phys. Rev. D",
    volume = "25",
    pages = "774",
    year = "1982"
}

@article{Fukugita:1986hr,
    author = "Fukugita, M. and Yanagida, T.",
    title = "{Baryogenesis Without Grand Unification}",
    reportNumber = "RIFP-641",
    doi = "10.1016/0370-2693(86)91126-3",
    journal = "Phys. Lett. B",
    volume = "174",
    pages = "45--47",
    year = "1986"
}

@article{Luty:1992un,
    author = "Luty, M. A.",
    title = "{Baryogenesis via leptogenesis}",
    doi = "10.1103/PhysRevD.45.455",
    journal = "Phys. Rev. D",
    volume = "45",
    pages = "455--465",
    year = "1992"
}

@article{Pilaftsis:1997jf,
    author = "Pilaftsis, Apostolos",
    title = "{CP violation and baryogenesis due to heavy Majorana neutrinos}",
    eprint = "hep-ph/9707235",
    archivePrefix = "arXiv",
    reportNumber = "MPI-PHT-97-30",
    doi = "10.1103/PhysRevD.56.5431",
    journal = "Phys. Rev. D",
    volume = "56",
    pages = "5431--5451",
    year = "1997"
}

@article{Ma:1998dx,
    author = "Ma, Ernest and Sarkar, Utpal",
    title = "{Neutrino masses and leptogenesis with heavy Higgs triplets}",
    eprint = "hep-ph/9802445",
    archivePrefix = "arXiv",
    reportNumber = "UCRHEP-T-209",
    doi = "10.1103/PhysRevLett.80.5716",
    journal = "Phys. Rev. Lett.",
    volume = "80",
    pages = "5716--5719",
    year = "1998"
}

@article{Hambye:2000ui,
    author = "Hambye, Thomas and Ma, Ernest and Sarkar, Utpal",
    title = "{Supersymmetric triplet Higgs model of neutrino masses and leptogenesis}",
    eprint = "hep-ph/0011192",
    archivePrefix = "arXiv",
    reportNumber = "CPT-00-P-4084, UCRHEP-T292",
    doi = "10.1016/S0550-3213(01)00109-2",
    journal = "Nucl. Phys. B",
    volume = "602",
    pages = "23--38",
    year = "2001"
}

@article{Hambye:2003ka,
    author = "Hambye, Thomas and Senjanovic, Goran",
    title = "{Consequences of triplet seesaw for leptogenesis}",
    eprint = "hep-ph/0307237",
    archivePrefix = "arXiv",
    doi = "10.1016/j.physletb.2003.11.061",
    journal = "Phys. Lett. B",
    volume = "582",
    pages = "73--81",
    year = "2004"
}

@article{Greene:1998nh,
    author = "Greene, Patrick B. and Kofman, Lev",
    title = "{Preheating of fermions}",
    eprint = "hep-ph/9807339",
    archivePrefix = "arXiv",
    reportNumber = "UH-IFA-98-44",
    doi = "10.1016/S0370-2693(99)00020-9",
    journal = "Phys. Lett. B",
    volume = "448",
    pages = "6--12",
    year = "1999"
}

@article{Lyth:1998xn,
    author = "Lyth, David H. and Riotto, Antonio",
    title = "{Particle physics models of inflation and the cosmological density perturbation}",
    eprint = "hep-ph/9807278",
    archivePrefix = "arXiv",
    reportNumber = "LANCS-TH-9720, FERMILAB-PUB-97-292-A, CERN-TH-97-383, OUTP-98-39-P",
    doi = "10.1016/S0370-1573(98)00128-8",
    journal = "Phys. Rept.",
    volume = "314",
    pages = "1--146",
    year = "1999"
}

@article{Guth:1980zm,
    author = "Guth, Alan H.",
    editor = "Fang, Li-Zhi and Ruffini, R.",
    title = "{The Inflationary Universe: A Possible Solution to the Horizon and Flatness Problems}",
    reportNumber = "SLAC-PUB-2576",
    doi = "10.1103/PhysRevD.23.347",
    journal = "Phys. Rev. D",
    volume = "23",
    pages = "347--356",
    year = "1981"
}

@article{Sato:1981qmu,
    author = "Sato, Katsuhiko",
    title = "{First-order phase transition of a vacuum and the expansion of the Universe}",
    doi = "10.1093/mnras/195.3.467",
    journal = "Mon. Not. Roy. Astron. Soc.",
    volume = "195",
    number = "3",
    pages = "467--479",
    year = "1981"
}

@article{Linde:1981mu,
    author = "Linde, Andrei D.",
    editor = "Fang, Li-Zhi and Ruffini, R.",
    title = "{A New Inflationary Universe Scenario: A Possible Solution of the Horizon, Flatness, Homogeneity, Isotropy and Primordial Monopole Problems}",
    reportNumber = "LEBEDEV-81-229",
    doi = "10.1016/0370-2693(82)91219-9",
    journal = "Phys. Lett. B",
    volume = "108",
    pages = "389--393",
    year = "1982"
}

@article{Linde:1983gd,
    author = "Linde, Andrei D.",
    title = "{Chaotic Inflation}",
    doi = "10.1016/0370-2693(83)90837-7",
    journal = "Phys. Lett. B",
    volume = "129",
    pages = "177--181",
    year = "1983"
}

@article{Riotto:2002yw,
    author = "Riotto, Antonio",
    editor = "Dvali, G. and Perez-Lorenzana, Abdel and Senjanovic, G. and Thompson, G. and Vissani, F.",
    title = "{Inflation and the theory of cosmological perturbations}",
    eprint = "hep-ph/0210162",
    archivePrefix = "arXiv",
    reportNumber = "DFPD-TH-02-22",
    journal = "ICTP Lect. Notes Ser.",
    volume = "14",
    pages = "317--413",
    year = "2003"
}

@article{Kinney:2003xf,
    author = "Kinney, William H.",
    editor = "Prosper, Harrison B. and Danilov, Michael",
    title = "{Cosmology, inflation, and the physics of nothing}",
    eprint = "astro-ph/0301448",
    archivePrefix = "arXiv",
    reportNumber = "CU-TP-1083",
    doi = "10.1007/978-94-010-0076-5_5",
    journal = "NATO Sci. Ser. II",
    volume = "123",
    pages = "189--243",
    year = "2003"
}

@inproceedings{Baumann:2009ds,
    author = "Baumann, Daniel",
    title = "{Inflation}",
    booktitle = "{Theoretical Advanced Study Institute in Elementary Particle Physics}: {Physics of the Large and the Small}",
    eprint = "0907.5424",
    archivePrefix = "arXiv",
    primaryClass = "hep-th",
    reportNumber = "TASI-2009",
    doi = "10.1142/9789814327183_0010",
    pages = "523--686",
    year = "2011"
}

@article{Kazanas:1980tx,
    author = "Kazanas, D.",
    title = "{Dynamics of the Universe and Spontaneous Symmetry Breaking}",
    doi = "10.1086/183361",
    journal = "Astrophys. J. Lett.",
    volume = "241",
    pages = "L59--L63",
    year = "1980"
}

@article{Olive:1989nu,
    author = "Olive, Keith A.",
    title = "{Inflation}",
    reportNumber = "UMN-TH-804-89",
    doi = "10.1016/0370-1573(90)90144-Q",
    journal = "Phys. Rept.",
    volume = "190",
    pages = "307--403",
    year = "1990"
}

@article{Shtanov:1994ce,
    author = "Shtanov, Y. and Traschen, Jennie H. and Brandenberger, Robert H.",
    title = "{Universe reheating after inflation}",
    eprint = "hep-ph/9407247",
    archivePrefix = "arXiv",
    reportNumber = "BROWN-HET-957",
    doi = "10.1103/PhysRevD.51.5438",
    journal = "Phys. Rev. D",
    volume = "51",
    pages = "5438--5455",
    year = "1995"
}

@article{Dolgov:1989us,
    author = "Dolgov, A. D. and Kirilova, D. P.",
    title = "{ON PARTICLE CREATION BY A TIME DEPENDENT SCALAR FIELD}",
    reportNumber = "JINR-E2-89-321",
    journal = "Sov. J. Nucl. Phys.",
    volume = "51",
    pages = "172--177",
    year = "1990"
}

@article{Traschen:1990sw,
    author = "Traschen, Jennie H. and Brandenberger, Robert H.",
    title = "{Particle Production During Out-of-equilibrium Phase Transitions}",
    reportNumber = "BROWN-HET-731",
    doi = "10.1103/PhysRevD.42.2491",
    journal = "Phys. Rev. D",
    volume = "42",
    pages = "2491--2504",
    year = "1990"
}

@article{Kofman:1994rk,
    author = "Kofman, Lev and Linde, Andrei D. and Starobinsky, Alexei A.",
    title = "{Reheating after inflation}",
    eprint = "hep-th/9405187",
    archivePrefix = "arXiv",
    reportNumber = "UH-IFA-94-35, SU-ITP-94-13, YITP-U-94-15",
    doi = "10.1103/PhysRevLett.73.3195",
    journal = "Phys. Rev. Lett.",
    volume = "73",
    pages = "3195--3198",
    year = "1994"
}

@article{Boyanovsky:1995ud,
    author = "Boyanovsky, D. and D'Attanasio, M. and de Vega, H. J. and Holman, R. and Lee, D. -S. and Singh, A.",
    title = "{Reheating the postinflationary universe}",
    eprint = "hep-ph/9505220",
    archivePrefix = "arXiv",
    reportNumber = "PITT-09-95, CMU-95-03, DOE-ER-40682-92, LPTHE-95-18, UPRF-95-420",
    month = "5",
    year = "1995"
}

@article{Yoshimura:1995gc,
    author = "Yoshimura, M.",
    title = "{Catastrophic particle production under periodic perturbation}",
    eprint = "hep-th/9506176",
    archivePrefix = "arXiv",
    reportNumber = "TU-484, TU-95-484",
    doi = "10.1143/PTP.94.873",
    journal = "Prog. Theor. Phys.",
    volume = "94",
    pages = "873--898",
    year = "1995"
}

@article{Giudice:2000ex,
    author = "Giudice, Gian Francesco and Kolb, Edward W. and Riotto, Antonio",
    title = "{Largest temperature of the radiation era and its cosmological implications}",
    eprint = "hep-ph/0005123",
    archivePrefix = "arXiv",
    reportNumber = "SNS-PH-00-05, FERMILAB-PUB-00-075-A, CERN-TH-2000-107",
    doi = "10.1103/PhysRevD.64.023508",
    journal = "Phys. Rev. D",
    volume = "64",
    pages = "023508",
    year = "2001"
}

@article{Kolb:2003ke,
    author = "Kolb, Edward W. and Notari, Alessio and Riotto, Antonio",
    title = "{On the Reheating Stage after Inflation}",
    eprint = "hep-ph/0307241",
    archivePrefix = "arXiv",
    reportNumber = "FERMILAB-PUB-03-212-A",
    doi = "10.1103/PhysRevD.68.123505",
    journal = "Phys. Rev. D",
    volume = "68",
    pages = "123505",
    year = "2003"
}

@book{Kolb:1990vq,
    author = "Kolb, Edward W. and Turner, Michael S.",
    title = "{The Early Universe}",
    reportNumber = "FERMILAB-BOOK-1990-01",
    doi = "10.1201/9780429492860",
    isbn = "978-0-429-49286-0, 978-0-201-62674-2",
    publisher = "Taylor and Francis",
    volume = "69",
    month = "5",
    year = "2019"
}

@article{Cook:2015vqa,
    author = "Cook, Jessica L. and Dimastrogiovanni, Emanuela and Easson, Damien A. and Krauss, Lawrence M.",
    title = "{Reheating predictions in single field inflation}",
    eprint = "1502.04673",
    archivePrefix = "arXiv",
    primaryClass = "astro-ph.CO",
    doi = "10.1088/1475-7516/2015/04/047",
    journal = "JCAP",
    volume = "04",
    pages = "047",
    year = "2015"
}

@article{Cai:2015soa,
    author = "Cai, Rong-Gen and Guo, Zong-Kuan and Wang, Shao-Jiang",
    title = "{Reheating phase diagram for single-field slow-roll inflationary models}",
    eprint = "1501.07743",
    archivePrefix = "arXiv",
    primaryClass = "gr-qc",
    doi = "10.1103/PhysRevD.92.063506",
    journal = "Phys. Rev. D",
    volume = "92",
    pages = "063506",
    year = "2015"
}

@article{Ichikawa:2005vw,
    author = "Ichikawa, Kazuhide and Kawasaki, Masahiro and Takahashi, Fuminobu",
    title = "{The Oscillation effects on thermalization of the neutrinos in the Universe with low reheating temperature}",
    eprint = "astro-ph/0505395",
    archivePrefix = "arXiv",
    doi = "10.1103/PhysRevD.72.043522",
    journal = "Phys. Rev. D",
    volume = "72",
    pages = "043522",
    year = "2005"
}

@article{Kawasaki:2000en,
    author = "Kawasaki, M. and Kohri, Kazunori and Sugiyama, Naoshi",
    title = "{MeV scale reheating temperature and thermalization of neutrino background}",
    eprint = "astro-ph/0002127",
    archivePrefix = "arXiv",
    doi = "10.1103/PhysRevD.62.023506",
    journal = "Phys. Rev. D",
    volume = "62",
    pages = "023506",
    year = "2000"
}

@article{Barbieri:2025moq,
    author = "Barbieri, Nicola and Brinckmann, Thejs and Gariazzo, Stefano and Lattanzi, Massimiliano and Pastor, Sergio and Pisanti, Ofelia",
    title = "{Current Constraints on Cosmological Scenarios with Very Low Reheating Temperatures}",
    eprint = "2501.01369",
    archivePrefix = "arXiv",
    primaryClass = "astro-ph.CO",
    doi = "10.1103/j5rj-dz1k",
    journal = "Phys. Rev. Lett.",
    volume = "135",
    number = "18",
    pages = "181003",
    year = "2025"
}

@article{Kallosh:2013lkr,
    author = "Kallosh, Renata and Linde, Andrei",
    title = "{Superconformal generalizations of the Starobinsky model}",
    eprint = "1306.3214",
    archivePrefix = "arXiv",
    primaryClass = "hep-th",
    doi = "10.1088/1475-7516/2013/06/028",
    journal = "JCAP",
    volume = "06",
    pages = "028",
    year = "2013"
}

@article{Kallosh:2013hoa,
    author = "Kallosh, Renata and Linde, Andrei",
    title = "{Universality Class in Conformal Inflation}",
    eprint = "1306.5220",
    archivePrefix = "arXiv",
    primaryClass = "hep-th",
    doi = "10.1088/1475-7516/2013/07/002",
    journal = "JCAP",
    volume = "07",
    pages = "002",
    year = "2013"
}

@article{Kallosh:2013yoa,
    author = "Kallosh, Renata and Linde, Andrei and Roest, Diederik",
    title = "{Superconformal Inflationary $\alpha$-Attractors}",
    eprint = "1311.0472",
    archivePrefix = "arXiv",
    primaryClass = "hep-th",
    doi = "10.1007/JHEP11(2013)198",
    journal = "JHEP",
    volume = "11",
    pages = "198",
    year = "2013"
}

@article{Kallosh:2013pby,
    author = "Kallosh, Renata and Linde, Andrei",
    title = "{Superconformal generalization of the chaotic inflation model $\frac{\lambda}{4} \phi^{4} - \frac{\xi}{2} \phi^{2}R$}",
    eprint = "1306.3211",
    archivePrefix = "arXiv",
    primaryClass = "hep-th",
    doi = "10.1088/1475-7516/2013/06/027",
    journal = "JCAP",
    volume = "06",
    pages = "027",
    year = "2013"
}

@article{Galante:2014ifa,
    author = "Galante, Mario and Kallosh, Renata and Linde, Andrei and Roest, Diederik",
    title = "{Unity of Cosmological Inflation Attractors}",
    eprint = "1412.3797",
    archivePrefix = "arXiv",
    primaryClass = "hep-th",
    doi = "10.1103/PhysRevLett.114.141302",
    journal = "Phys. Rev. Lett.",
    volume = "114",
    number = "14",
    pages = "141302",
    year = "2015"
}

@article{Giudice:1999fb,
    author = "Giudice, G. F. and Peloso, M. and Riotto, A. and Tkachev, I.",
    title = "{Production of massive fermions at preheating and leptogenesis}",
    eprint = "hep-ph/9905242",
    archivePrefix = "arXiv",
    reportNumber = "CERN-TH-99-117",
    doi = "10.1088/1126-6708/1999/08/014",
    journal = "JHEP",
    volume = "08",
    pages = "014",
    year = "1999"
}

@article{Peloso:2000hy,
    author = "Peloso, Marco and Sorbo, Lorenzo",
    title = "{Preheating of massive fermions after inflation: Analytical results}",
    eprint = "hep-ph/0003045",
    archivePrefix = "arXiv",
    doi = "10.1088/1126-6708/2000/05/016",
    journal = "JHEP",
    volume = "05",
    pages = "016",
    year = "2000"
}

@article{Khlebnikov:1997di,
    author = "Khlebnikov, S. Y. and Tkachev, I. I.",
    title = "{Relic gravitational waves produced after preheating}",
    eprint = "hep-ph/9701423",
    archivePrefix = "arXiv",
    reportNumber = "PURD-TH-97-02, OSU-TA-01-97",
    doi = "10.1103/PhysRevD.56.653",
    journal = "Phys. Rev. D",
    volume = "56",
    pages = "653--660",
    year = "1997"
}

@article{Easther:2006gt,
    author = "Easther, Richard and Lim, Eugene A.",
    title = "{Stochastic gravitational wave production after inflation}",
    eprint = "astro-ph/0601617",
    archivePrefix = "arXiv",
    doi = "10.1088/1475-7516/2006/04/010",
    journal = "JCAP",
    volume = "04",
    pages = "010",
    year = "2006"
}

@article{Easther:2006vd,
    author = "Easther, Richard and Giblin, Jr., John T. and Lim, Eugene A.",
    title = "{Gravitational Wave Production At The End Of Inflation}",
    eprint = "astro-ph/0612294",
    archivePrefix = "arXiv",
    doi = "10.1103/PhysRevLett.99.221301",
    journal = "Phys. Rev. Lett.",
    volume = "99",
    pages = "221301",
    year = "2007"
}

@article{Garcia-Bellido:2007nns,
    author = "Garcia-Bellido, Juan and Figueroa, Daniel G.",
    title = "{A stochastic background of gravitational waves from hybrid preheating}",
    eprint = "astro-ph/0701014",
    archivePrefix = "arXiv",
    reportNumber = "IFT-UAM-CSIC-06-46",
    doi = "10.1103/PhysRevLett.98.061302",
    journal = "Phys. Rev. Lett.",
    volume = "98",
    pages = "061302",
    year = "2007"
}

@article{Garcia-Bellido:2007fiu,
    author = "Garcia-Bellido, Juan and Figueroa, Daniel G. and Sastre, Alfonso",
    title = "{A Gravitational Wave Background from Reheating after Hybrid Inflation}",
    eprint = "0707.0839",
    archivePrefix = "arXiv",
    primaryClass = "hep-ph",
    reportNumber = "IFT-UAM-CSIC-07-38",
    doi = "10.1103/PhysRevD.77.043517",
    journal = "Phys. Rev. D",
    volume = "77",
    pages = "043517",
    year = "2008"
}

@article{Dufaux:2007pt,
    author = "Dufaux, Jean Francois and Bergman, Amanda and Felder, Gary N. and Kofman, Lev and Uzan, Jean-Philippe",
    title = "{Theory and Numerics of Gravitational Waves from Preheating after Inflation}",
    eprint = "0707.0875",
    archivePrefix = "arXiv",
    primaryClass = "astro-ph",
    doi = "10.1103/PhysRevD.76.123517",
    journal = "Phys. Rev. D",
    volume = "76",
    pages = "123517",
    year = "2007"
}

@article{Dufaux:2008dn,
    author = "Dufaux, Jean-Francois and Felder, Gary and Kofman, Lev and Navros, Olga",
    title = "{Gravity Waves from Tachyonic Preheating after Hybrid Inflation}",
    eprint = "0812.2917",
    archivePrefix = "arXiv",
    primaryClass = "astro-ph",
    reportNumber = "FTUAM-08-25, IFT-UAM-CSIC-08-90",
    doi = "10.1088/1475-7516/2009/03/001",
    journal = "JCAP",
    volume = "03",
    pages = "001",
    year = "2009"
}

@article{Figueroa:2011ye,
    author = "Figueroa, Daniel G. and Garcia-Bellido, Juan and Rajantie, Arttu",
    title = "{On the Transverse-Traceless Projection in Lattice Simulations of Gravitational Wave Production}",
    eprint = "1110.0337",
    archivePrefix = "arXiv",
    primaryClass = "astro-ph.CO",
    reportNumber = "HIP-2011-26, IFT-UAM-CSIC-11-54, IMPERIAL-TP-2011-AR-01",
    doi = "10.1088/1475-7516/2011/11/015",
    journal = "JCAP",
    volume = "11",
    pages = "015",
    year = "2011"
}

@article{Bethke:2013aba,
    author = "Bethke, Laura and Figueroa, Daniel G. and Rajantie, Arttu",
    title = "{Anisotropies in the Gravitational Wave Background from Preheating}",
    eprint = "1304.2657",
    archivePrefix = "arXiv",
    primaryClass = "astro-ph.CO",
    reportNumber = "IMPERIAL-TP-2013-LB-1",
    doi = "10.1103/PhysRevLett.111.011301",
    journal = "Phys. Rev. Lett.",
    volume = "111",
    number = "1",
    pages = "011301",
    year = "2013"
}

@article{Bethke:2013vca,
    author = "Bethke, Laura and Figueroa, Daniel G. and Rajantie, Arttu",
    title = "{On the Anisotropy of the Gravitational Wave Background from Massless Preheating}",
    eprint = "1309.1148",
    archivePrefix = "arXiv",
    primaryClass = "astro-ph.CO",
    doi = "10.1088/1475-7516/2014/06/047",
    journal = "JCAP",
    volume = "06",
    pages = "047",
    year = "2014"
}

@article{Figueroa:2017vfa,
    author = "Figueroa, Daniel G. and Torrenti, Francisco",
    title = "{Gravitational wave production from preheating: parameter dependence}",
    eprint = "1707.04533",
    archivePrefix = "arXiv",
    primaryClass = "astro-ph.CO",
    reportNumber = "CERN-TH-2017-152, IFT-UAM-CSIC-17-069",
    doi = "10.1088/1475-7516/2017/10/057",
    journal = "JCAP",
    volume = "10",
    pages = "057",
    year = "2017"
}

@article{Garcia:2021iag,
    author = "Garcia, Marcos A. G. and Kaneta, Kunio and Mambrini, Yann and Olive, Keith A. and Verner, Sarunas",
    title = "{Freeze-in from preheating}",
    eprint = "2109.13280",
    archivePrefix = "arXiv",
    primaryClass = "hep-ph",
    reportNumber = "UMN-TH-4101/21, FTPI-MINN-21/19, CERN-TH-2021-121",
    doi = "10.1088/1475-7516/2022/03/016",
    journal = "JCAP",
    volume = "03",
    number = "03",
    pages = "016",
    year = "2022"
}

@article{Garcia:2022vwm,
    author = "Garcia, Marcos A. G. and Pierre, Mathias and Verner, Sarunas",
    title = "{Scalar dark matter production from preheating and structure formation constraints}",
    eprint = "2206.08940",
    archivePrefix = "arXiv",
    primaryClass = "hep-ph",
    reportNumber = "DESY-22-104",
    doi = "10.1103/PhysRevD.107.043530",
    journal = "Phys. Rev. D",
    volume = "107",
    number = "4",
    pages = "043530",
    year = "2023"
}

@article{Abbott:1982hn,
    author = "Abbott, L. F. and Farhi, Edward and Wise, Mark B.",
    title = "{Particle Production in the New Inflationary Cosmology}",
    reportNumber = "MIT-CTP-983",
    doi = "10.1016/0370-2693(82)90867-X",
    journal = "Phys. Lett. B",
    volume = "117",
    pages = "29",
    year = "1982"
}

@article{Albrecht:1982mp,
    author = "Albrecht, Andreas and Steinhardt, Paul J. and Turner, Michael S. and Wilczek, Frank",
    title = "{Reheating an Inflationary Universe}",
    reportNumber = "UPR-0189T, EFI-82-09-CHICAGO",
    doi = "10.1103/PhysRevLett.48.1437",
    journal = "Phys. Rev. Lett.",
    volume = "48",
    pages = "1437",
    year = "1982"
}

@article{Martin:2010kz,
    author = "Martin, Jerome and Ringeval, Christophe",
    title = "{First CMB Constraints on the Inflationary Reheating Temperature}",
    eprint = "1004.5525",
    archivePrefix = "arXiv",
    primaryClass = "astro-ph.CO",
    doi = "10.1103/PhysRevD.82.023511",
    journal = "Phys. Rev. D",
    volume = "82",
    pages = "023511",
    year = "2010"
}

@article{Adshead:2010mc,
    author = "Adshead, Peter and Easther, Richard and Pritchard, Jonathan and Loeb, Abraham",
    title = "{Inflation and the Scale Dependent Spectral Index: Prospects and Strategies}",
    eprint = "1007.3748",
    archivePrefix = "arXiv",
    primaryClass = "astro-ph.CO",
    doi = "10.1088/1475-7516/2011/02/021",
    journal = "JCAP",
    volume = "02",
    pages = "021",
    year = "2011"
}

@article{Mielczarek:2010ag,
    author = "Mielczarek, Jakub",
    title = "{Reheating temperature from the CMB}",
    eprint = "1009.2359",
    archivePrefix = "arXiv",
    primaryClass = "astro-ph.CO",
    doi = "10.1103/PhysRevD.83.023502",
    journal = "Phys. Rev. D",
    volume = "83",
    pages = "023502",
    year = "2011"
}

@article{Dai:2014jja,
    author = "Dai, Liang and Kamionkowski, Marc and Wang, Junpu",
    title = "{Reheating constraints to inflationary models}",
    eprint = "1404.6704",
    archivePrefix = "arXiv",
    primaryClass = "astro-ph.CO",
    doi = "10.1103/PhysRevLett.113.041302",
    journal = "Phys. Rev. Lett.",
    volume = "113",
    pages = "041302",
    year = "2014"
}

@article{Martin:2014nya,
    author = "Martin, Jerome and Ringeval, Christophe and Vennin, Vincent",
    title = "{Observing Inflationary Reheating}",
    eprint = "1410.7958",
    archivePrefix = "arXiv",
    primaryClass = "astro-ph.CO",
    doi = "10.1103/PhysRevLett.114.081303",
    journal = "Phys. Rev. Lett.",
    volume = "114",
    number = "8",
    pages = "081303",
    year = "2015"
}

@article{Drewes:2017fmn,
    author = "Drewes, Marco and Kang, Jin U and Mun, Ui Ri",
    title = "{CMB constraints on the inflaton couplings and reheating temperature in $\alpha$-attractor inflation}",
    eprint = "1708.01197",
    archivePrefix = "arXiv",
    primaryClass = "astro-ph.CO",
    reportNumber = "TUM-HEP-1093-17",
    doi = "10.1007/JHEP11(2017)072",
    journal = "JHEP",
    volume = "11",
    pages = "072",
    year = "2017"
}

@article{Drewes:2022nhu,
    author = "Drewes, Marco and Ming, Lei",
    title = "{Connecting Cosmic Inflation to Particle Physics with LiteBIRD, CMB-S4, EUCLID, and SKA}",
    eprint = "2208.07609",
    archivePrefix = "arXiv",
    primaryClass = "hep-ph",
    doi = "10.1103/PhysRevLett.133.031001",
    journal = "Phys. Rev. Lett.",
    volume = "133",
    number = "3",
    pages = "031001",
    year = "2024"
}

@article{Podolsky:2005bw,
    author = "Podolsky, Dmitry I. and Felder, Gary N. and Kofman, Lev and Peloso, Marco",
    title = "{Equation of state and beginning of thermalization after preheating}",
    eprint = "hep-ph/0507096",
    archivePrefix = "arXiv",
    reportNumber = "UMN-TH-2407-05",
    doi = "10.1103/PhysRevD.73.023501",
    journal = "Phys. Rev. D",
    volume = "73",
    pages = "023501",
    year = "2006"
}

@article{DiMarco:2021xzk,
    author = "Di Marco, Alessandro and Pradisi, Gianfranco",
    title = "{Variable inflaton equation-of-state and reheating}",
    eprint = "2102.00326",
    archivePrefix = "arXiv",
    primaryClass = "gr-qc",
    doi = "10.1142/S0217751X21500950",
    journal = "Int. J. Mod. Phys. A",
    volume = "36",
    number = "15",
    pages = "2150095",
    year = "2021"
}

@phdthesis{Qutub:2017wnf,
    author = "Qutub, Saleh Sadaka O.",
    title = "{Dark Matter Production after Inflation and Constraints}",
    doi = "10.17635/lancaster/thesis/67",
    school = "Lancaster U.",
    year = "2017"
}

@article{Nakayama:2018ptw,
    author = "Nakayama, Kazunori and Tang, Yong",
    title = "{Stochastic Gravitational Waves from Particle Origin}",
    eprint = "1810.04975",
    archivePrefix = "arXiv",
    primaryClass = "hep-ph",
    reportNumber = "UT-18-20",
    doi = "10.1016/j.physletb.2018.11.023",
    journal = "Phys. Lett. B",
    volume = "788",
    pages = "341--346",
    year = "2019"
}

@article{Ghoshal:2022kqp,
    author = "Ghoshal, Anish and Samanta, Rome and White, Graham",
    title = "{Bremsstrahlung high-frequency gravitational wave signatures of high-scale nonthermal leptogenesis}",
    eprint = "2211.10433",
    archivePrefix = "arXiv",
    primaryClass = "hep-ph",
    doi = "10.1103/PhysRevD.108.035019",
    journal = "Phys. Rev. D",
    volume = "108",
    number = "3",
    pages = "035019",
    year = "2023"
}

@article{Datta:2024tne,
    author = "Datta, Arghyajit and Sil, Arunansu",
    title = "{Probing Leptogenesis through Gravitational Waves}",
    eprint = "2410.01900",
    archivePrefix = "arXiv",
    primaryClass = "hep-ph",
    month = "10",
    year = "2024"
}

@article{Xu:2024fjl,
    author = "Xu, Yong",
    title = "{Ultra-high frequency gravitational waves from scattering, Bremsstrahlung and decay during reheating}",
    eprint = "2407.03256",
    archivePrefix = "arXiv",
    primaryClass = "hep-ph",
    reportNumber = "MITP-24-058",
    doi = "10.1007/JHEP10(2024)174",
    journal = "JHEP",
    volume = "10",
    pages = "174",
    year = "2024"
}

@article{Choi:2024acs,
    author = "Choi, Ki-Young and Lkhagvadorj, Erdenebulgan and Mahapatra, Satyabrata",
    title = "{Gravitational wave sourced by decay of massive particle from primordial black hole evaporation}",
    eprint = "2403.15269",
    archivePrefix = "arXiv",
    primaryClass = "hep-ph",
    doi = "10.1088/1475-7516/2024/07/064",
    journal = "JCAP",
    volume = "07",
    pages = "064",
    year = "2024"
}

@article{Watkins:1991zt,
    author = "Watkins, Richard and Widrow, Lawrence M.",
    title = "{Aspects of reheating in first order inflation}",
    reportNumber = "FERMILAB-PUB-91-164-A",
    doi = "10.1016/0550-3213(92)90362-F",
    journal = "Nucl. Phys. B",
    volume = "374",
    pages = "446--468",
    year = "1992"
}

@article{Konstandin:2011ds,
    author = "Konstandin, Thomas and Servant, Geraldine",
    title = "{Natural Cold Baryogenesis from Strongly Interacting Electroweak Symmetry Breaking}",
    eprint = "1104.4793",
    archivePrefix = "arXiv",
    primaryClass = "hep-ph",
    doi = "10.1088/1475-7516/2011/07/024",
    journal = "JCAP",
    volume = "07",
    pages = "024",
    year = "2011"
}

@article{Falkowski:2012fb,
    author = "Falkowski, Adam and No, Jose M.",
    title = "{Non-thermal Dark Matter Production from the Electroweak Phase Transition: Multi-TeV WIMPs and 'Baby-Zillas'}",
    eprint = "1211.5615",
    archivePrefix = "arXiv",
    primaryClass = "hep-ph",
    reportNumber = "ULB-TH-12-18, LPT-12-113",
    doi = "10.1007/JHEP02(2013)034",
    journal = "JHEP",
    volume = "02",
    pages = "034",
    year = "2013"
}

@article{Dasgupta:2022isg,
    author = "Dasgupta, Arnab and Dev, P. S. Bhupal and Ghoshal, Anish and Mazumdar, Anupam",
    title = "{Gravitational wave pathway to testable leptogenesis}",
    eprint = "2206.07032",
    archivePrefix = "arXiv",
    primaryClass = "hep-ph",
    doi = "10.1103/PhysRevD.106.075027",
    journal = "Phys. Rev. D",
    volume = "106",
    number = "7",
    pages = "075027",
    year = "2022"
}

@article{Cataldi:2024pgt,
    author = "Cataldi, Martina and Shakya, Bibhushan",
    title = "{Leptogenesis via bubble collisions}",
    eprint = "2407.16747",
    archivePrefix = "arXiv",
    primaryClass = "hep-ph",
    reportNumber = "DESY-24-110",
    doi = "10.1088/1475-7516/2024/11/047",
    journal = "JCAP",
    volume = "11",
    pages = "047",
    year = "2024"
}

@article{Baker:2019nia,
    author = "Baker, John and others",
    title = "{The Laser Interferometer Space Antenna: Unveiling the Millihertz Gravitational Wave Sky}",
    eprint = "1907.06482",
    archivePrefix = "arXiv",
    primaryClass = "astro-ph.IM",
    reportNumber = "FERMILAB-PUB-19-436-A",
    month = "7",
    year = "2019"
}

@article{Corbin:2005ny,
    author = "Corbin, Vincent and Cornish, Neil J.",
    title = "{Detecting the cosmic gravitational wave background with the big bang observer}",
    eprint = "gr-qc/0512039",
    archivePrefix = "arXiv",
    doi = "10.1088/0264-9381/23/7/014",
    journal = "Class. Quant. Grav.",
    volume = "23",
    pages = "2435--2446",
    year = "2006"
}

@article{Crowder:2005nr,
    author = "Crowder, Jeff and Cornish, Neil J.",
    title = "{Beyond LISA: Exploring future gravitational wave missions}",
    eprint = "gr-qc/0506015",
    archivePrefix = "arXiv",
    doi = "10.1103/PhysRevD.72.083005",
    journal = "Phys. Rev. D",
    volume = "72",
    pages = "083005",
    year = "2005"
}

@article{Harry:2006fi,
    author = "Harry, G. M. and Fritschel, P. and Shaddock, D. A. and Folkner, W. and Phinney, E. S.",
    title = "{Laser interferometry for the big bang observer}",
    doi = "10.1088/0264-9381/23/15/008",
    journal = "Class. Quant. Grav.",
    volume = "23",
    pages = "4887--4894",
    year = "2006",
    note = "[Erratum: Class.Quant.Grav. 23, 7361 (2006)]"
}

@article{Seto:2001qf,
    author = "Seto, Naoki and Kawamura, Seiji and Nakamura, Takashi",
    title = "{Possibility of direct measurement of the acceleration of the universe using 0.1-Hz band laser interferometer gravitational wave antenna in space}",
    eprint = "astro-ph/0108011",
    archivePrefix = "arXiv",
    doi = "10.1103/PhysRevLett.87.221103",
    journal = "Phys. Rev. Lett.",
    volume = "87",
    pages = "221103",
    year = "2001"
}

@article{Kawamura:2020pcg,
    author = "Kawamura, Seiji and others",
    title = "{Current status of space gravitational wave antenna DECIGO and B-DECIGO}",
    eprint = "2006.13545",
    archivePrefix = "arXiv",
    primaryClass = "gr-qc",
    doi = "10.1093/ptep/ptab019",
    journal = "PTEP",
    volume = "2021",
    number = "5",
    pages = "05A105",
    year = "2021"
}

@article{Kawamura:2006up,
    author = "Kawamura, S. and others",
    editor = "Mio, N.",
    title = "{The Japanese space gravitational wave antenna DECIGO}",
    doi = "10.1088/0264-9381/23/8/S17",
    journal = "Class. Quant. Grav.",
    volume = "23",
    pages = "S125--S132",
    year = "2006"
}

@article{Greene:1997fu,
    author = "Greene, Patrick B. and Kofman, Lev and Linde, Andrei D. and Starobinsky, Alexei A.",
    title = "{Structure of resonance in preheating after inflation}",
    eprint = "hep-ph/9705347",
    archivePrefix = "arXiv",
    reportNumber = "SU-ITP-97-19, IFA-97-29",
    doi = "10.1103/PhysRevD.56.6175",
    journal = "Phys. Rev. D",
    volume = "56",
    pages = "6175--6192",
    year = "1997"
}

@article{Ben-Dayan:2019gll,
    author = "Ben-Dayan, Ido and Keating, Brian and Leon, David and Wolfson, Ira",
    title = "{Constraints on scalar and tensor spectra from $N_{eff}$}",
    eprint = "1903.11843",
    archivePrefix = "arXiv",
    primaryClass = "astro-ph.CO",
    doi = "10.1088/1475-7516/2019/06/007",
    journal = "JCAP",
    volume = "06",
    pages = "007",
    year = "2019"
}

@article{EUCLID:2011zbd,
    author = "Laureijs, R. and others",
    collaboration = "EUCLID",
    title = "{Euclid Definition Study Report}",
    eprint = "1110.3193",
    archivePrefix = "arXiv",
    primaryClass = "astro-ph.CO",
    reportNumber = "ESA-SRE(2011)12",
    month = "10",
    year = "2011"
}

@article{COrE:2011bfs,
    author = "Bouchet, F. R. and others",
    collaboration = "COrE",
    title = "{COrE (Cosmic Origins Explorer) A White Paper}",
    eprint = "1102.2181",
    archivePrefix = "arXiv",
    primaryClass = "astro-ph.CO",
    month = "2",
    year = "2011"
}

@article{Lozanov:2016hid,
    author = "Lozanov, Kaloian D. and Amin, Mustafa A.",
    title = "{Equation of State and Duration to Radiation Domination after Inflation}",
    eprint = "1608.01213",
    archivePrefix = "arXiv",
    primaryClass = "astro-ph.CO",
    doi = "10.1103/PhysRevLett.119.061301",
    journal = "Phys. Rev. Lett.",
    volume = "119",
    number = "6",
    pages = "061301",
    year = "2017"
}

@article{Murayama:2025thw,
    author = {Murayama, Hitoshi and Noether, Bea and Sch{\"u}tte-Engel, Jan},
    title = "{Observing leptogenesis in action with gravitational waves}",
    eprint = "2506.15772",
    archivePrefix = "arXiv",
    primaryClass = "hep-ph",
    reportNumber = "RIKEN-iTHEMS-Report-25",
    doi = "10.1088/1475-7516/2025/12/027",
    journal = "JCAP",
    volume = "12",
    pages = "027",
    year = "2025"
}

@article{Datta:2025wfh,
    author = "Datta, Arghyajit and Khalil, Shaaban and Mandal, Rajat Kumar and Sil, Arunansu",
    title = "{Probing right handed neutrino assisted reheating with gravitational waves and leptogenesis}",
    eprint = "2507.09728",
    archivePrefix = "arXiv",
    primaryClass = "hep-ph",
    doi = "10.1088/1475-7516/2026/02/061",
    journal = "JCAP",
    volume = "02",
    pages = "061",
    year = "2026"
}

@article{Inui:2024wgj,
    author = "Inui, Ryoto and Mikura, Yusuke and Yokoyama, Shuichiro",
    title = "{Gravitational waves from graviton bremsstrahlung with kination phase}",
    eprint = "2408.10786",
    archivePrefix = "arXiv",
    primaryClass = "astro-ph.CO",
    doi = "10.1103/PhysRevD.111.043511",
    journal = "Phys. Rev. D",
    volume = "111",
    number = "4",
    pages = "043511",
    year = "2025"
}

@article{Hu:2024bha,
    author = "Hu, Wei-Yu and Nakayama, Kazunori and Takhistov, Volodymyr and Tang, Yong",
    title = "{Dual gravitational wave signatures of instant preheating}",
    eprint = "2409.06483",
    archivePrefix = "arXiv",
    primaryClass = "astro-ph.CO",
    reportNumber = "TU-1240, KEK-QUP-2024-0020, KEK-TH-2650, KEK-Cosmo-0356",
    doi = "10.1088/1475-7516/2025/01/029",
    journal = "JCAP",
    volume = "01",
    pages = "029",
    year = "2025"
}

@article{Hu:2024awd,
    author = "Hu, Weiyu and Nakayama, Kazunori and Takhistov, Volodymyr and Tang, Yong",
    title = "{Gravitational wave probe of Planck-scale physics after inflation}",
    eprint = "2403.13882",
    archivePrefix = "arXiv",
    primaryClass = "hep-ph",
    reportNumber = "KEK-QUP-2024-0006, TU-1226, KEK-TH-2607, KEK-Cosmo-0341, IPMU24-0007",
    doi = "10.1016/j.physletb.2024.138958",
    journal = "Phys. Lett. B",
    volume = "856",
    pages = "138958",
    year = "2024"
}

@article{Konar:2025iuk,
    author = "Konar, Partha and Show, Sudipta",
    title = "{Unraveling Freeze-in Dark matter through the echoes of gravitational waves}",
    eprint = "2506.08106",
    archivePrefix = "arXiv",
    primaryClass = "hep-ph",
    month = "6",
    year = "2025"
}

@article{Ai:2025fqw,
    author = "Ai, Wen-Yuan",
    title = "{High-frequency gravitational waves from a first-order phase transitions}",
    eprint = "2508.02794",
    archivePrefix = "arXiv",
    primaryClass = "hep-ph",
    doi = "10.1103/4v2z-fqsx",
    journal = "Phys. Rev. D",
    volume = "113",
    number = "5",
    pages = "056007",
    year = "2026"
}

@article{Haque:2023zhb,
    author = "Haque, Md Riajul and Maity, Debaprasad and Mondal, Rajesh",
    title = "{Gravitational neutrino reheating}",
    eprint = "2311.07684",
    archivePrefix = "arXiv",
    primaryClass = "hep-ph",
    doi = "10.1103/PhysRevD.109.063543",
    journal = "Phys. Rev. D",
    volume = "109",
    number = "6",
    pages = "063543",
    year = "2024"
}

@article{Haque:2024zdq,
    author = "Haque, Md Riajul and Maity, Debaprasad and Mondal, Rajesh",
    title = "{Thermal and nonthermal dark matters with gravitational neutrino reheating}",
    eprint = "2408.12450",
    archivePrefix = "arXiv",
    primaryClass = "hep-ph",
    doi = "10.1103/PhysRevD.111.063546",
    journal = "Phys. Rev. D",
    volume = "111",
    number = "6",
    pages = "063546",
    year = "2025"
}

@article{Cataldi:2025nac,
    author = {Cataldi, Martina and M{\"u}{\"u}rsepp, Kristjan and Vanvlasselaer, Miguel},
    title = "{CP-violation in production of heavy neutrinos from bubble collisions}",
    eprint = "2506.12123",
    archivePrefix = "arXiv",
    primaryClass = "hep-ph",
    reportNumber = "DESY-25-082",
    doi = "10.1007/JHEP01(2026)058",
    journal = "JHEP",
    volume = "01",
    pages = "058",
    year = "2026"
}

@article{Herman:2022fau,
    author = "Herman, Nicolas and Lehoucq, L{\'e}onard and F{\'{u}}zfa, Andr{\'e}",
    title = "{Electromagnetic antennas for the resonant detection of the stochastic gravitational wave background}",
    eprint = "2203.15668",
    archivePrefix = "arXiv",
    primaryClass = "gr-qc",
    doi = "10.1103/PhysRevD.108.124009",
    journal = "Phys. Rev. D",
    volume = "108",
    number = "12",
    pages = "124009",
    year = "2023"
}

@article{Antoniadis:2025pfa,
    author = "Antoniadis, Ignatios and Ellis, John and Ke, Wenqi and Nanopoulos, Dimitri V. and Olive, Keith A.",
    title = "{How accidental was inflation?}",
    eprint = "2504.12283",
    archivePrefix = "arXiv",
    primaryClass = "hep-ph",
    reportNumber = "UMN-TH-4418/25, FTPI-MINN-25/03, KCL-PH-TH/2025-09, CERN-TH-2025-076",
    doi = "10.1088/1475-7516/2025/08/090",
    journal = "JCAP",
    volume = "08",
    pages = "090",
    year = "2025"
}

@article{Biswas:2025adi,
    author = "Biswas, Anirban and Ganguly, Sougata and Nanda, Dibyendu and Sahoo, Sujit Kumar",
    title = "{Viability of post-inflationary freeze-in with precision cosmology}",
    eprint = "2505.13624",
    archivePrefix = "arXiv",
    primaryClass = "hep-ph",
    reportNumber = "CTPU-PTC-25-19",
    month = "5",
    year = "2025"
}

@article{Berlin:2021txa,
    author = {Berlin, Asher and Blas, Diego and Tito D'Agnolo, Raffaele and Ellis, Sebastian A. R. and Harnik, Roni and Kahn, Yonatan and Sch{\"u}tte-Engel, Jan},
    title = "{Detecting high-frequency gravitational waves with microwave cavities}",
    eprint = "2112.11465",
    archivePrefix = "arXiv",
    primaryClass = "hep-ph",
    reportNumber = "FERMILAB-PUB-21-724-SQMS-T",
    doi = "10.1103/PhysRevD.105.116011",
    journal = "Phys. Rev. D",
    volume = "105",
    number = "11",
    pages = "116011",
    year = "2022"
}

@article{Yeh:2022heq,
    author = "Yeh, Tsung-Han and Shelton, Jessie and Olive, Keith A. and Fields, Brian D.",
    title = "{Probing physics beyond the standard model: limits from BBN and the CMB independently and combined}",
    eprint = "2207.13133",
    archivePrefix = "arXiv",
    primaryClass = "astro-ph.CO",
    reportNumber = "UMN-TH-4125/22, FTPI-MINN-22/16",
    doi = "10.1088/1475-7516/2022/10/046",
    journal = "JCAP",
    volume = "10",
    pages = "046",
    year = "2022"
}

\end{document}